\documentclass[12pt]{article}
\usepackage{a4wide,amssymb,cite,slashed}
\pdfoutput=1

\usepackage{a4wide,amssymb,cite,graphicx}
\usepackage{epsfig}
\usepackage[usenames,dvipsnames]{color}
\usepackage{slashed}
\newcommand{\be}{\begin{equation}}
\newcommand{\ee}{\end{equation}}
\newcommand{\bea}{\begin{eqnarray}}
\newcommand{\eea}{\end{eqnarray}}

\def\circa#1{\,\raise.3ex\hbox{$#1$\kern-.75em\lower1ex\hbox{$\sim$}}\,}

\begin{document}

\begin{titlepage}

\begin{centering}
\vspace{1cm}
{\Large {\bf Dark Matter Direct Detection from new interactions \vspace{.2cm} \\ in models with spin-two mediators  }} \\

\vspace{1.5cm}

{\bf  A. Carrillo-Monteverde$^{1,\sharp}$, Yoo-Jin Kang$^{2,3,*}$, Hyun Min Lee$^{2,\dagger}$ , \vspace{0.15cm}\\ Myeonghun Park$^{4,\clubsuit}$, and Veronica Sanz$^{1,\spadesuit}$}
\vspace{.5cm}

{\it $^1$Department of Physics and Astronomy, University of Sussex, Brighton BN1 9QH, UK.}\vspace{0.15cm}
\\
{\it $^2$Department of Physics, Chung-Ang University, Seoul 06974, Korea.} \vspace{0.15cm}
\\
{\it $^3$Center for Theoretical Physics of the Universe,   \\ Institute for Basic Science, 
 Daejeon, 34051, Korea. }\vspace{0.15cm}
\\
{\it $^4$Institute of Convergence Fundamental Studies and School of Liberal Arts, \\
Seoul National University of Science and Technology, Seoul 01811, Korea.}

\end{centering}
\vspace{2cm}

\begin{abstract}
We consider models where a massive spin-two resonance acts as the mediator between Dark Matter (DM) and the SM particles through the energy-momentum tensor. We examine the effective theory for fermion, vector and scalar DM generated in these models and find novel types of DM-SM interaction never considered before. We identify the effective interactions between DM and the SM quarks when the mediator is integrated out, and match them to the gravitational form factors relevant for spin-independent DM-nucleon scattering.  We also discuss the interplay between DM relic density conditions, direct detection bounds and collider searches for the spin-two mediator. 
\end{abstract}

\vspace{3cm}
  \begin{flushleft}
    $^\sharp$Email: A.carrillo-Monteverde@sussex.ac.uk \\    
    $^*$Email: yoojinkang91@gmail.com \\
    $^\dagger$Email: hminlee@cau.ac.kr \\
     $^\clubsuit$Email: ishaed@gmail.com   \\
     $^\spadesuit$Email: v.sanz@sussex.ac.uk
  \end{flushleft}

\end{titlepage}

\section{Introduction}

Weakly Interacting Massive Particles (WIMPs) have been well-motivated candidates for dark matter (DM) and being searched for in direct detection, cosmic rays as well as collider experiments. In particular, weak-scale interactions of DM have been strongly constrained by direct detection experiments \cite{xenon1t,panda,cdms,lux}, although dependent on astrophysical inputs such as the DM velocity distribution and the local DM density.

The interactions between dark matter and SM particles can be well described in terms of effective operators at the scale of momentum transfer, typically below $100\,{\rm MeV}$. However, the validity of the same effective interactions is in question at the freeze-out of the annihilation of dark matter in the early Universe or at collider experiments such as the Large Hadron Collider (LHC). For these reasons, the mediator particles, that are responsible for the effective interactions, have been introduced in simplified models for dark matter \cite{dmforum}. In this case, integrating out the mediator particles lead to the effective operators for direct detection and mediator particles can be directly produced in the early Universe or at the LHC. Most of the proposed mediator models, however, have been focused upon spin-0 and spin-1 mediators, for which the correlated effective interactions are limited \cite{fitz,fitz2,cirelli,zurek} and only the effective operators lower than dimension-8 have been taken into account until now \cite{zupan}.

We consider a massive spin-2 resonance as  the mediator that couples to dark matter with arbitrary spin and the SM particles through the energy-momentum tensor, as considered in Refs.~\cite{GMDM1,GMDM2,diphoton,cascade,mawatari,rizzo}.  After the spin-2 mediator is integrated out, we identify the  effective interactions between dark matter and the SM quarks up to dimension-8 and match them to the gravitational form factors for nucleons beyond a zero momentum transfer.  
Focusing on fermion or scalar dark matter, we show how the non-relativistic operators between dark matter and nucleons are correlated. 
We also discuss how in our model of spin-two mediators there is an interplay between relic density conditions, current direct detection bounds and searches at the LHC of dijet resonances produced in association with a jet or a photon.

The paper is organized as follows. 
We first present the general setup for the effective interactions between dark matter and the SM quarks in the presence of a massive spin-2 mediator and consider the matching conditions for them to the gravitational form factors for nucleons. 
Then, we derive the scattering amplitudes of fermion or scalar dark matter with nucleons and compute the corresponding differential scattering events rates for DM direct detection experiments.
Next we impose the bounds from the correct relic density condition and the direct detection experiments for the parameter space of our model.
Finally, conclusions are drawn. 
There are three appendices reviewing the differential scattering event rates, the scattering amplitudes between dark matter and nucleon, and the nucleon matrix elements for twist-2 operators at zero momentum transfer.

\section{Spin-2 mediator and dark matter}

We present the interactions between dark matter and the SM particles which are mediated by a massive spin-2 mediator.  Focusing on the spin-2 mediator couplings to quarks in the SM for direct detection experiments, we identify the effective interactions for the SM quarks and their counterpart gravitational form factors for nucleons.

\subsection{Effective interactions between dark matter and quarks}

We introduce the couplings of a massive spin-2 mediator to the SM particles and dark matter, through the energy-momentum tensor, as follows \cite{GMDM1,GMDM2,diphoton,cascade},
\bea
{\cal L}_{\rm int}= -\frac{c_{\rm SM}}{\Lambda} {\cal G}^{\mu\nu} T^{\rm SM}_{\mu\nu} -\frac{c_{\rm DM}}{\Lambda} {\cal G}^{\mu\nu} T^{\rm DM}_{\mu\nu}.
\eea
In this case, the mediator couplings for the SM particles can vary, depending on the location in the extra dimension \cite{GMDM1,GMDM2}.
Then, the tree-level scattering amplitude between DM and SM particles through the spin-2 mediator is given by
\bea
{\cal M}=-\frac{c_{\rm DM} c_{\rm SM} }{\Lambda^2} \frac{i}{q^2-m^2_G}\,T^{\rm DM}_{\mu\nu}(q){\cal P}^{\mu\nu,\alpha\beta}(q) T^{\rm SM}_{\alpha\beta}(-q)  \label{ampl}
\eea
where $q$ is the 4-momentum transfer between dark matter and the SM particles and the tensor structure for the massive spin-2 propagator is given by
\bea
{\cal P}_{\mu\nu,\alpha\beta}(q)=\frac{1}{2}\Big(G_{\mu\alpha}G_{\nu\beta}+G_{\nu\alpha} G_{\mu\beta}- \frac{2}{3} G_{\mu\nu} G_{\alpha\beta}\Big)
\eea
with
\bea
G_{\mu\nu}\equiv \eta_{\mu\nu}- \frac{q_\mu q_\nu}{m^2_G}. 
\eea
We note that the sum of the spin-2 mediator polarizations is given by
\bea
\sum_s \epsilon_{\mu\nu}(q,s) \epsilon_{\alpha\beta}(q,s)= P_{\mu\nu,\alpha\beta}(q). 
\eea
The tensor $P_{\mu\nu,\alpha\beta}$ satisfies traceless and transverse conditions for on-shell spin-2 mediator, such as $\eta^{\alpha\beta} P_{\mu\nu,\alpha\beta}(q)=0$ and $q^\alpha P_{\mu\nu,\alpha\beta}(q)=0$ \cite{GMDM1}.

Due to energy-momentum conservation, $q_\mu T^{\mu\nu}=0$, we can replace $G_{\mu\nu}$ in the scattering amplitude (\ref{ampl}) by $\eta_{\mu\nu}$. Then, the scattering amplitude is divided into trace and traceless parts of energy-momentum tensor, as follows,
\bea
{\cal M}&=&\frac{ic_{\rm DM} c_{\rm SM}}{2m^2_G\Lambda^2}\, \bigg(2 T^{\rm DM}_{\mu\nu} T^{{\rm SM},\mu\nu} -\frac{2}{3}T^{\rm DM} T^{\rm SM}\bigg)  \nonumber \\
&=&\frac{ic_{\rm DM} c_{\rm SM}}{2m^2_G\Lambda^2}\, \bigg(2{\tilde T}^{\rm DM}_{\mu\nu} {\tilde T}^{{\rm SM},\mu\nu} -\frac{1}{6}{T}^{\rm DM} { T}^{\rm SM}\bigg)
\eea
where ${\tilde T}^{\rm SM(DM)}_{\mu\nu}$ is the traceless part of energy-momentum tensor given by ${\tilde T}^{\rm SM(DM)}_{\mu\nu}=T^{\rm SM(DM)}_{\mu\nu}-\frac{1}{4}\eta_{\mu\nu} {T}^{\rm SM(DM)}$ with ${T}^{\rm SM(DM)}$ being the trace of energy-momentum tensor.

The energy momentum tensor for the SM quarks denoted by $\psi$ \cite{GMDM1} is, in momentum space, 
\bea
T^\psi_{\mu\nu}= -\frac{1}{4} {\bar u}_\psi(p_2) \Big(\gamma_\mu (p_{1\nu}+p_{2\nu})+\gamma_\nu (p_{1\mu}+p_{2\mu})-2\eta_{\mu\nu}(\slashed{p}_1+\slashed{p}_2-2m_\psi) \Big)u_\psi(p_1) \label{em-quark}
\eea
where $u_\psi(p)$ is the Dirac spinor for $\psi$. Here, the SM fermion is incoming into the vertex with momentum $p_1$ and is outgoing from the vertex with momentum $p_2$.
Then, the trace of the energy-momentum tensor for $\psi$ is given by
\bea
T^\psi=-\frac{1}{4} {\bar u}_\psi(p_2)\Big(-6(\slashed{p}_1+\slashed{p}_2)+16m_\psi\Big)u_\psi(p_1).
\eea
Moreover, the traceless part of the energy-momentum tensor  or the twist-2 operator for $\psi$ is given by
\bea
{\tilde T}^\psi_{\mu\nu}= -\frac{1}{4} {\bar u}_\psi(p_2)\Big(\gamma_\mu (p_{1\nu}+p_{2\nu})+\gamma_\nu (p_{1\mu}+p_{2\mu})-\frac{1}{2}\eta_{\mu\nu}(\slashed{p}_1+\slashed{p}_2) \Big)u_\psi(p_1).
\eea

\subsection{Gravitational form factors for nucleons}

For the trace part of the energy-momentum tensor, we now match to the nuclear matrix elements  as
\bea
\langle N(p_2)|T^\psi|N(p_1)\rangle=- F_S(q^2) m_N {\bar u}_N(p_2)u_N(p_1) \label{nucl-trace}
\eea
where $F_S(q^2)$ is the form factor for the scalar current, given at $q=0$ by $F_S(0)= f^N_{T\psi}$  as in eq.~(\ref{scalar-0}) in Appendix C. The momentum expansion of the scalar form factor $F_S(q^2)$ is given in Ref.~\cite{zupan}.

On the other hand, the energy-momentum tensor (\ref{em-quark}) is matched to the nuclear matrix elements, as follows \cite{Gform,holo},
\bea
\langle N(p_2)|T^\psi_{\mu\nu}|N(p_1)\rangle
&=&{\bar u}_N(p_2) \Big(A(q^2)\gamma_{(\mu} p_{\nu)}+B(q^2)\,\frac{1}{2m_N}\, i p_{(\mu}\sigma_{\nu)\lambda}q^\lambda \nonumber \\
&&+C(q^2)\frac{1}{m_N} (q_\mu q_\nu-\eta_{\mu\nu}q^2 ) \Big) u_N(p_1) \nonumber \\
&=&{\bar u}_N(p_2) \Big(\frac{2}{m_N}\,A(q^2)p_\mu p_\nu+(A(q^2)+B(q^2))\,\frac{1}{2m_N}\, i p_{(\mu}\sigma_{\nu)\lambda}q^\lambda \nonumber \\
&&+C(q^2)\frac{1}{m_N} (q_\mu q_\nu-\eta_{\mu\nu}q^2 ) \Big) u_N(p_1) \label{gform}
\eea
where $A(q^2), B(q^2), C(q^2)$ are the gravitational form factors, and $p_\mu= \frac{1}{2}(p_1+p_2)$ and $q=p_2-p_1$ and $(\,)$ means the symmetrization of indices.  Here we have used the Gordon identity in the second equality and we note that the second term is the anomalous gravitational magnetic moment operator.
One can check that the above form factors, $A(q^2), B(q^2), C(q^2)$, are the only ones that are consistent with Lorentz invariance, $q_\mu T^{\psi,\mu\nu}=0$, and $CP$ symmetry \cite{Gform,holo}.

By using the energy-momentum tensor for on-shell nucleons,
\bea
T^N_{\mu\nu}&=&-\frac{1}{2} {\bar u}_N(p_2) \gamma_{(\mu} p_{\nu)}  u_N(p_1) \nonumber \\
&=&-\frac{1}{2} {\bar u}_N(p_2) \Big(\frac{2}{m_N}\,p_\mu p_\nu +\frac{1}{2m_N}\,i p_{(\mu}\sigma_{\nu)\lambda}q^\lambda  \Big),
\eea 
where use is made of the Gordon identity in the second equality, we can rewrite the above nuclear matrix elements in eq.~(\ref{gform}) as
\bea
\langle N(p_2)|T^\psi_{\mu\nu}|N(p_1)\rangle&=&-2(A(q^2)+B(q^2)) T^N_{\mu\nu} \nonumber \\
&&+\frac{1}{m_N}\, {\bar u}_N(p_2)\Big(-2B(q^2)p_\mu p_\nu+C(q^2) (q_\mu q_\nu-\eta_{\mu\nu}q^2 )  \Big) u_N(p_1)   \label{nucl-em}
\eea 
where  $T^N_{\mu\nu}$ is the energy-momentum tensor for nucleons.
Then, using eq.~(\ref{nucl-em}) and its trace, we obtain the nuclear matrix elements for the twist-2 operator as 
\bea
\langle N(p_2)|{\tilde T}^\psi_{\mu\nu}|N(p_1)\rangle&=&-2(A(q^2)+B(q^2)) {\tilde T}^N_{\mu\nu} +\frac{1}{m_N} \,{\bar u}_N(p_2) \bigg[-2B(q^2) \Big(p_\mu p_\nu-\frac{1}{4}g_{\mu\nu} p^2 \Big)  \nonumber \\
&&+ C(q^2) \Big(q_\mu q_\nu - \frac{1}{4} \eta_{\mu\nu} q^2 \Big) \bigg] u_N(p_1). \label{twist2-q}
\eea
  
For nucleons with zero momentum transfer, the nuclear matrix elements for the twist-2 quark operator \cite{drees,hisano,solon} becomes
\bea
\langle N(p)| {\tilde T}^\psi_{\mu\nu}|N(p)\rangle= -\frac{1}{m_N}\,F_T(0) \Big(p_\mu p_\nu-\frac{1}{4} m^2_N g_{\mu\nu}\Big) {\bar u}_N(p)u_N(p)  \label{twist2-0}
\eea  
where the form factor for the twist-2 quark operator is given by $F_T(0)\equiv -2 A(0)=\psi(2)+{\bar \psi}(2)$, as in eq.~(\ref{twistquark}) with eq.~(\ref{pdf2nd}) and the following discussion in Appendix C. 
Therefore, $B(q^2), C(q^2)$ remain unfixed due to lack of the extra information on the form factors for a nonzero momentum transfer. 

In a holographic description of QCD with the hard-wall or soft-wall model in a five-dimensional Anti-de Sitter (AdS) spacetime \cite{holo}, the gravitational form factors can be described by the three-point correlation function between the zero-mode graviton and a fermion in the bulk on the boundary of the AdS spacetime. In this case, the relation between the gravitational form factors is given by $A(q^2)\neq 0$ and $B(q^2)=C(q^2)=0$ even for a nonzero momentum transfer \cite{holo}.
In this case, we can extend the matching of the quark operators in eq.~(\ref{twist2-q}) to the nucleon form factors with a general momentum transfer as follows,  
\bea
\langle N(p_2)|{\tilde T}^\psi_{\mu\nu}|N(p_1)\rangle=F_T(q^2) {\tilde T}^N_{\mu\nu} \label{nucl-tensor}
\eea
with $F_T(q^2)\equiv -2A(q^2)$. Thus, the matching just relates the energy-momentum tensors for quarks and nucleons by the overall form factor, $F_T(q^2)$. This form factor has been explicitly computed in a holographic set-up with a {\it soft wall} model. This would be the dual of theories with a conformal UV limit and a softly broken symmetry at low energies. The breaking is then spontaneous and the features of the soft wall allow for switching on an non-zero expectation value for an operator with finite canonical dimension or a non-AdS background mass of the dual fields, see e.g. ~\cite{Hirn:2006wg}.  

\begin{figure}[t!]
  \begin{center}
      \includegraphics[width=0.5\textwidth]{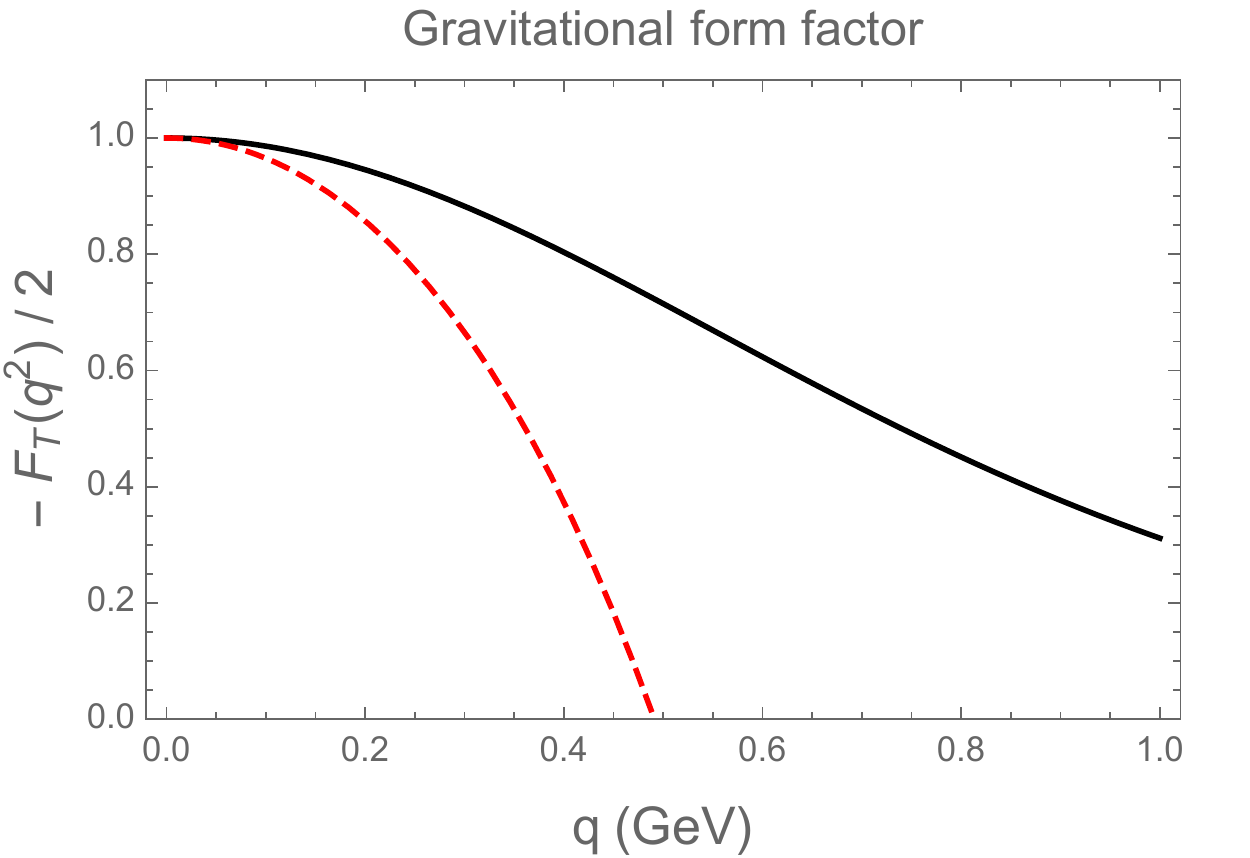}   
   \end{center}
       \caption{Gravitational form factor $-F_T(q^2)/2$ as a function of the momentum transfer $q$ in GeV. The black line corresponds to the exact expression obtained in Ref.~\cite{holo} and the same input parameters. The red-dashed line corresponds to the quadratic approximation. }
  \label{fig:form}
\end{figure}

In this context, Ref.~\cite{holo} finds an explicit form of the form factor, which decreases with $q^2$ as shown in Fig.~\ref{fig:form}, and admits an expansion near $q^2\simeq 0$ as follows
\bea
F_T(q^2) \simeq -2 ( 1 - q^2/(0.55 \textrm{ GeV})^2  \ldots ) .
\eea
Henceforth, for simplicity, we assume that this is the case and also take $F_T(q^2)\approx F_T(0)$. Further improvements in the analysis could be done by simultaneously expanding the form factor coupling $F_T(q^2)$ and following the standard procedure of non-relativistic expansion described in the next Section. 

We also remark the twist-2 gluon operator, which could be also relevant for direct detection in the presence of the gluon couplings of the spin-2 mediator.
The twist-2 gluon operator and its matching to nucleon matrix elements at a zero momentum transfer is also included in Appendix C. In this case, we may also identify gravitational form factors  for the twist-2 gluon operator at a nonzero momentum transfer by extending the result in eq.~(\ref{twistgluon}).
  But, in the current work we focus on the SM quark operators  and postpone the details on the twist-2 gluon operators for future work.

\section{Effective operators for DM-nucleon scattering}

In this section, we discuss the effective Lagrangian for the elastic scattering between dark matter and nucleons due to the spin-2 mediator.  We focus on the cases with fermion and scalar dark matter in this section and present the further details for the calculation of scattering amplitudes including the case for vector dark matter  in appendix B.

\subsection{Fermion dark matter}

The energy-momentum tensor for a fermion DM $\chi$  is, in momentum space,
\bea
T^\chi_{\mu\nu}= -\frac{1}{4} {\bar u}_\chi(k_2)\Big(\gamma_\mu (k_{1\nu}+k_{2\nu})+\gamma_\nu (k_{1\mu}+k_{2\mu})-2\eta_{\mu\nu}(\slashed{k}_1+\slashed{k}_2-2m_\chi) \Big)u_\chi(k_1)
\eea
where the fermion DM is incoming into the vertex with momentum $k_1$ and is outgoing from the vertex with momentum $k_2$.
Then, the trace of the energy-momentum tensor is given by
\bea
T^\chi=-\frac{1}{4} {\bar u}_\chi(k_2)\Big(-6(\slashed{k}_1+\slashed{k}_2)+16m_\chi\Big)u_\chi(k_1). \label{fdm-scalar}
\eea
Therefore, the traceless part of the energy-momentum tensor is given by
\bea
{\tilde T}^\chi_{\mu\nu}= -\frac{1}{4}{\bar u}_\chi(k_2)\Big(\gamma_\mu (k_{1\nu}+k_{2\nu})+\gamma_\nu (k_{1\mu}+k_{2\mu})-\frac{1}{2}\eta_{\mu\nu}(\slashed{k}_1+\slashed{k}_2) \Big) u_\chi(k_1). \label{fdm-tensor}
\eea

We consider the elastic scattering between the DM fermion and the nucleon, $\chi(k_1)+N(p_1)\rightarrow \chi(k_2)+N(p_2)$. 
Using the nucleon matrix elements in eq.~(\ref{nucl-tensor}), we get 
\bea
{\tilde T}^\chi_{\mu\nu}\langle N(p_2)|{\tilde T}^{\rm  \psi,\mu\nu}|N(p_1)\rangle=F_T(q^2){\tilde T}^\chi_{\mu\nu} {\tilde T}^{N,\mu\nu}
\eea
where $F_T(q^2)\approx F_T(0)$.
From eqs.~(\ref{fdm-scalar}) and (\ref{nucl-trace}), the effective interactions for trace parts are
\bea
T^\chi \langle N(p_2)| T^\psi |N(p_1)\rangle
= m_\chi m_N F_S ({\bar u}_\chi(k_2)u_\chi(k_1))({\bar u}_N(p_2)u_N(p_1)). \label{trace}
\eea
Thus, the trace parts contain only scalar-scalar operators. 
Therefore, from eqs.~(\ref{tracefree}) and (\ref{trace}), 
we get the scattering amplitude between fermion dark matter and nucleon as follows,
 \bea
 {\cal M}_\chi= \frac{ic_\chi c_\psi}{2m^2_G \Lambda^2} \, \langle N(p_2)|\left(2{\tilde T}^\chi_{\mu\nu} {\tilde T}^{\psi,\mu\nu} -\frac{1}{6} {T}^\chi {T}^\psi \right)|N(p_1)\rangle, \label{scatt-fermion0}
 \eea
 the detailed form of which is given in eq.~(\ref{scatt-fermion}).

There appear five effective interactions between fermion dark matter and nucleon due to the spin-2 mediator, each of which matches with non-relativistic operators \cite{fitz} as in Table~\ref{tabla:sencilla}.
Here, we note that the non-relativistic nucleon operators are given \cite{fitz} by
\bea
{\cal O}^{\rm NR}_1&=&1, \quad 
{\cal O}^{\rm NR}_2=(v^{\bot})^2, \quad 
{\cal O}^{\rm NR}_3= i{\vec s}_N\cdot \Big(\frac{{\vec q}}{m_N}\times {\vec v}^\bot\Big),  \nonumber \\
{\cal O}^{\rm NR}_4& =& {\vec s}_\chi \cdot {\vec s}_N, \quad {\cal O}^{\rm NR}_5=i{\vec s}_\chi\cdot \Big(\frac{{\vec q}}{m_N}\times {\vec v}^\bot\Big) \quad {\cal O}^{\rm NR}_6 = \Big({\vec s}_\chi\cdot \frac{{\vec q}}{m_N}\Big)\Big({\vec s}_N\cdot \frac{{\vec q}}{m_N}\Big).
\eea
Here, ${\vec s}_\chi, {\vec s}_N$ are the spins of dark matter and nucleon, respectively, and $i{\vec q}, {\vec v}^\bot$  are Galilean, Hermitian invariants \cite{fitz}, meaning the momentum transfer and the relative velocity between dark matter and nucleon after scattering, respectively, and the latter is related to the initial relative velocity ${\vec v}$ and the momentum transfer by ${\vec v}^\bot={\vec v}+\frac{{\vec q}}{2\mu_N}$ with $\mu_N$ being the reduced mass of the DM-nucleon system and it satisfies ${\vec v}^\bot\cdot{\vec q}=0$.
We note that ${\cal O}^{\rm NR}_{1,2,5}$ give rise to only the spin-independent elastic scattering while ${\cal O}^{\rm NR}_{3,4,6}$ lead to the spin-dependent elastic scattering. 
All the appearing operators are $T$-even and $P$-even.

\begin{table}[htbp]
\begin{center}
\begin{tabular}{|c|c|c|}
\hline
 & $\mathcal{O}_i$   & $\sum_k \mathcal{O}^{\rm NR}_k$   \\ 
\hline
\hline 
F & $ \left(\bar{\chi}\chi\right) \left(\bar{N}N\right) $   & $ 4 m_{\chi} m_N \mathcal{O}_1^{NR}$\\ 
\hline
F & $ \left(\bar{\chi}\chi\right) \left(K_{\nu}\bar{N}i \sigma^{\nu\lambda}q_{\lambda} N\right) $  & $ 4 m_{\chi}^2 {\vec q}^2 \mathcal{O}_1^{NR}-16 m_{\chi}^2 m_N^2 \mathcal{O}_3^{NR}  $\\ 
\hline
F & $ \left(P_{\mu}\bar{\chi}i\sigma^{\mu\rho}q_{\rho}\chi\right) \left(\bar{N}N\right) $   & $ -4 m_N^2 {\vec q}^2 \mathcal{O}_1^{NR}+16 m_{\chi} m_N^3 \mathcal{O}_5^{NR} $\\ 
\hline
F & $ \left(\bar{\chi}i\sigma^{\mu\rho}q_{\rho}\chi\right) \left(\bar{N}i \sigma^{\nu\lambda}q_{\lambda} N\right)$  & $ 16 m_{\chi} m_N ({\vec q}^2 \mathcal{O}_4^{NR}- m_N^2\mathcal{O}_6^{NR}) $\\ 
\hline
F & $ \left(P_{\mu}\bar{\chi}i\sigma^{\mu\rho}q_{\rho}\chi\right)\left(K_{\nu}\bar{N}i \sigma^{\nu\lambda}q_{\lambda} N\right) $ &  $ -4 m_{\chi} m_N ({\vec q}^2 \mathcal{O}_1^{NR}-4 m_N^2 \mathcal{O}_3^{NR})$\\ 
 & &   $\times({\vec q}^2 \mathcal{O}_1^{NR}-4 m_{\chi}m_N \mathcal{O}_5^{NR}) $\\ 
\hline
\hline
S & $(S^* S)(\bar{N} N)$ & $2 m_N \mathcal{O}_1^{NR}$\\
\hline
S & $i(S^*\partial_\mu S-S\partial_\mu S^*)(\bar{N}\gamma^\mu N)$ &$4 m_S m_N  \mathcal{O}_1^{NR}$\\
\hline
\hline
V & ${\bar N}N$ & $2m_N f(\epsilon_1,\epsilon^*_2){\cal O}^{\rm NR}_1$\\
\hline
V & $\epsilon^{\alpha}_{1,2} {\bar N}i\sigma_{\alpha\lambda} q^\lambda N$ & $4i m_N^2\Big( {\vec s}_N\cdot({\vec \epsilon}_{1,2}\times \frac{\vec q}{m_N})\Big) $\\
\hline
V & $k_{1,2\nu} {\bar N}i\sigma^{\nu\lambda}q_\lambda N$  & $ m_\chi\Big({\vec q}^2{\cal O}^{\rm NR}_1-4 m_N^2 {\cal O}^{\rm NR}_3\Big)$\\
\hline
\end{tabular}
\caption{Effective operators for fermion (F), scalar (S) and vector (V) dark matter.}
\label{tabla:sencilla}
\end{center}
\end{table}

For the coherent scattering of nucleons, the momentum transfer is given by $|{\vec q}|\leq\sqrt{m_T E_R}\lesssim 100\,{\rm MeV}$ where $m_T$ is the target nucleus mass and $E_R$ is the recoil energy of the nucleus.
Therefore, for WIMP dark matter, using the results in eq.~(\ref{feff}), we obtain the approximate interaction Lagrangian between fermion dark matter and nucleons, as follows,
\bea
\mathcal{L}_{\chi,\rm eff}&\approx& \frac{c_\chi c_\psi m_\chi^2 m_N^2}{2m_G^2 \Lambda^2}\bigg[ \bigg\{6 F_T\Big(1+\frac{{\vec q}^2}{3m^2_N}+\frac{{\vec q}^2}{3m^2_\chi} \Big)  -\frac{2}{3}F_S  \bigg\}  {\mathcal{O}}_1^{\rm NR}-8F_T   {\mathcal{O}}_3^{\rm NR} -\frac{4{\vec q}^2}{m_\chi m_N}\, F_T {\mathcal{O}}_4^{\rm NR}   \nonumber \\
&&\quad -\frac{8m_N}{m_\chi} F_T   \Big(1+\frac{{\vec q}^2}{8m^2_N} \Big) {\mathcal{O}}_5^{\rm NR} +\frac{4m_N}{m_\chi}\, F_T  {\mathcal{O}}_6^{\rm NR}+\frac{4m_N}{m_\chi}\, F_T {\mathcal{O}}_3^{\rm NR} {\mathcal{O}}_5^{\rm NR}  \bigg]. \label{feff3}
\eea
The scalar operator ${\mathcal{O}}_1^{\rm NR}$ determines dominantly the total cross section for spin-independent elastic scattering as will be discussed in Section 5.4.
Other operators also contribute to the differential event rates, in particular, for a large momentum transfer or  recoil energy, but the momentum-dependent terms show up less than $1\%$.

\subsection{Scalar dark matter}

The energy-momentum tensor for  a scalar DM $S$ is, in momentum space,
\bea
T^S_{\mu\nu}= - \Big( m^2_S \eta_{\mu\nu}+C_{\mu\nu,\alpha\beta} k^\alpha_1 k^\beta_2 \Big)
\eea
where 
\bea
C_{\mu\nu,\alpha\beta}\equiv \eta_{\mu\alpha}\eta_{\nu\beta}+\eta_{\nu\alpha} \eta_{\mu\beta}-\eta_{\mu\nu}\eta_{\alpha\beta}
\eea
and the scalar DM is incoming into the vertex with momentum $k_1$ and is outgoing from the vertex with momentum $k_2$. 
Then, the trace of the energy-momentum tensor is given by
\bea
T^S=-\Big(4m^2_S - 2 (k_1\cdot k_2)\Big).
\eea
Therefore, the traceless part of the energy-momentum tensor is given by
\bea
{\tilde T}^S_{\mu\nu}= -\Big(k_{1\mu}k_{2\nu}+k_{2\mu}k_{1\nu}-\frac{1}{2}\eta_{\mu\nu}(k_1\cdot k_2) \Big).   \label{sdm-tensor}
\eea

We consider the elastic scattering between the DM scalar and the nucleon, $S(k_1)+N(p_1)\rightarrow S(k_2)+N(p_2)$. 
Then, similarly to the case of fermion dark matter,  from eq.~(\ref{nucl-tensor}), we get
\bea
{\tilde T}^S_{\mu\nu}\langle N(p_2)|{\tilde T}^{\rm \psi,{\mu\nu}}|N(p_1)\rangle= F_T(q^2) {\tilde T}^S_{\mu\nu} {\tilde T}^{N,\mu\nu}
\eea
with $F_T(q^2)\approx F_T(0)$.
On the other hand, the effective interactions for trace parts are
\bea
4T^S \langle N(p_2)|T^\psi|N(p_1)\rangle =8m_N F_S (2m^2_S-k_1\cdot k_2)\, ({\bar u}_N(p_2)u_N(p_1)).  \label{traceS}
\eea

Consequently, combining eqs.~(\ref{tracefreeS}) and (\ref{traceS}), 
 we get the scattering amplitude between scalar dark matter and nucleon, as follows,
 \bea
 {\cal M}_S= \frac{ic_S c_\psi}{2m^2_G \Lambda^2} \,\langle N(p_2)| \left(2{\tilde T}^S_{\mu\nu} {\tilde T}^{\psi,\mu\nu} -\frac{1}{6} {T}^S{ T}^\psi \right)|N(p_1)\rangle,   \label{scatt-scalar0}
\eea
the detailed form of which is given in eq.~(\ref{scatt-scalar}).

Therefore, for scalar dark matter, there appear two effective nucleon interactions at the relativistic level,  due to the spin-2 mediator, each of which matches to the non-relativistic nucleon operator \cite{cirelli} as shown in Table~\ref{tabla:sencilla}.  Then, using eq.~(\ref{seff}) with eq.~(\ref{approx}), we get the above effective Lagrangian as follows,
\bea
\mathcal{L}_{S, \rm eff}&=&\frac{c_S c_\psi m_S^2 m_N^2}{2m_G^2\Lambda^2}\bigg[ F_T  \bigg( 6  -\frac{{\vec q}^2}{m^2_S} \bigg) -\frac{2}{3}F_S\Big(1 -\frac{{\vec q}^2}{2m^2_S}\Big)  \bigg] {\cal O}_1^{\rm NR}. \label{seff2}
\eea
Thus, we find that the effective operators for scalar dark matter are reduced   to the scalar operator ${\mathcal{O}}_1^{\rm NR}$ at the non-relativistic level. We note that the above effective Lagrangian shows that ${\vec q}^2$ terms are highly suppressed by dark matter mass, as compared to the case for fermion dark matter in eq.~(\ref{feff3}) where ${\vec q}^2$ terms are suppressed just by nucleon mass.  
But, the total cross section for spin-independent elastic scattering is determined mainly by the scalar operator ${\mathcal{O}}_1^{\rm NR}$ as for fermion dark matter, which will become manifest from the same form of the total cross section for spin-dependent elastic scattering in Section 5.4.

\section{Differential scattering event rates with spin-2 mediator}

\begin{figure}[t!]
  \begin{center}
      \includegraphics[width=0.45\textwidth]{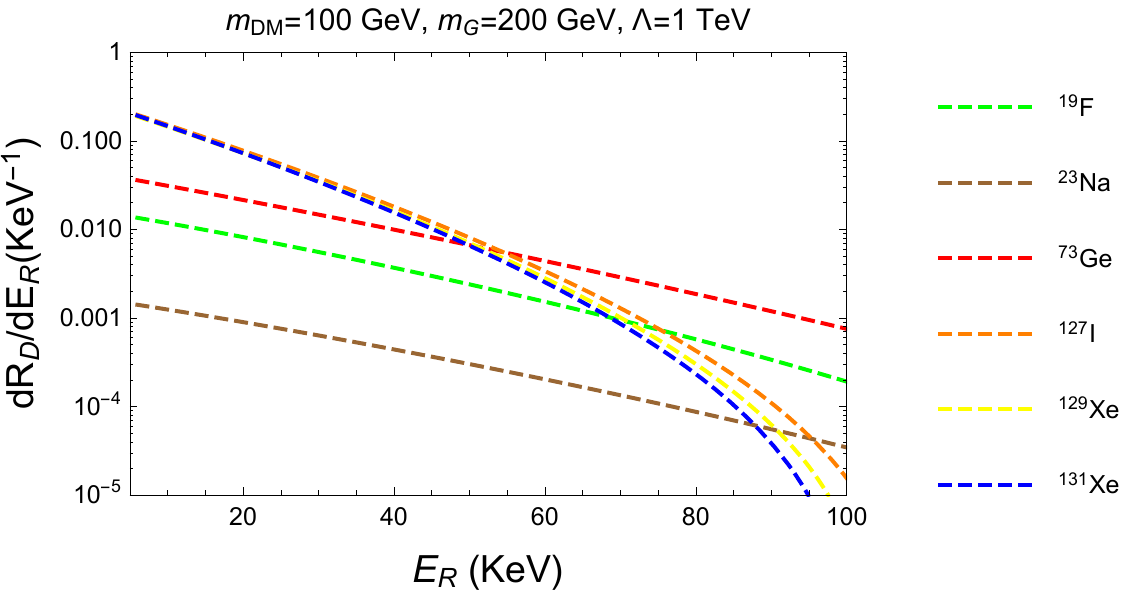} 
 		\includegraphics[width=0.45\textwidth]{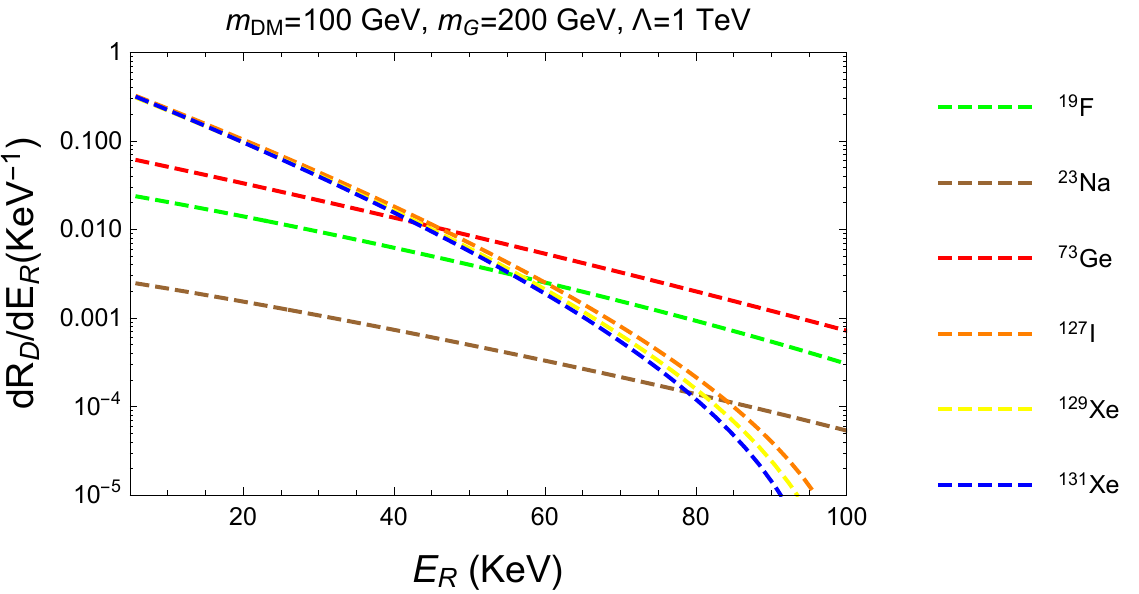}      
      \\
      \vspace{0.15cm}
          \      \includegraphics[width=0.45\textwidth]{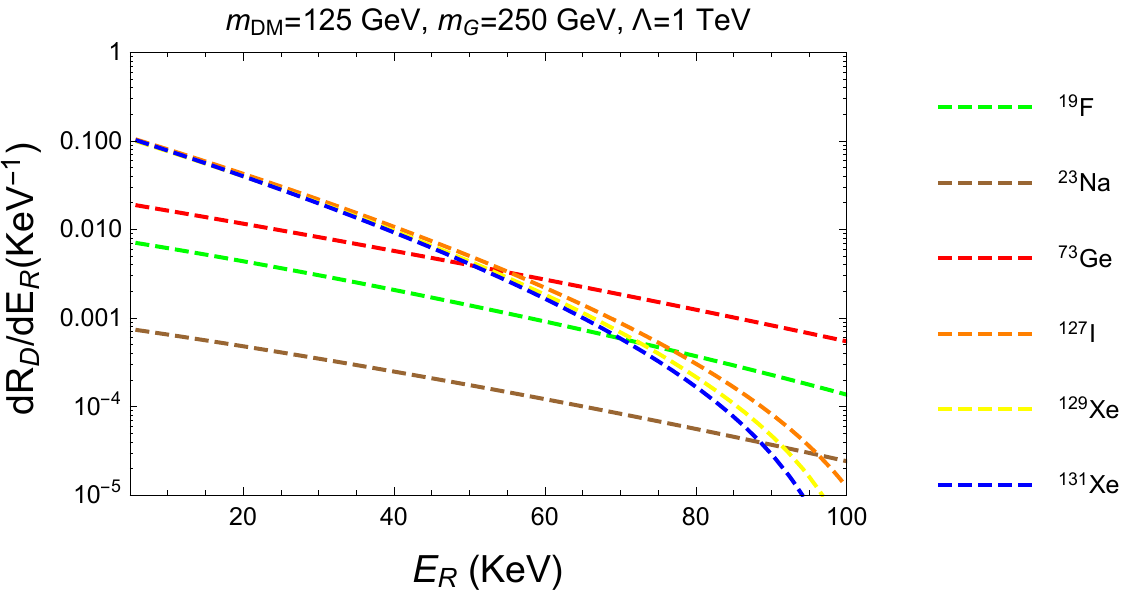} 
 		\includegraphics[width=0.45\textwidth]{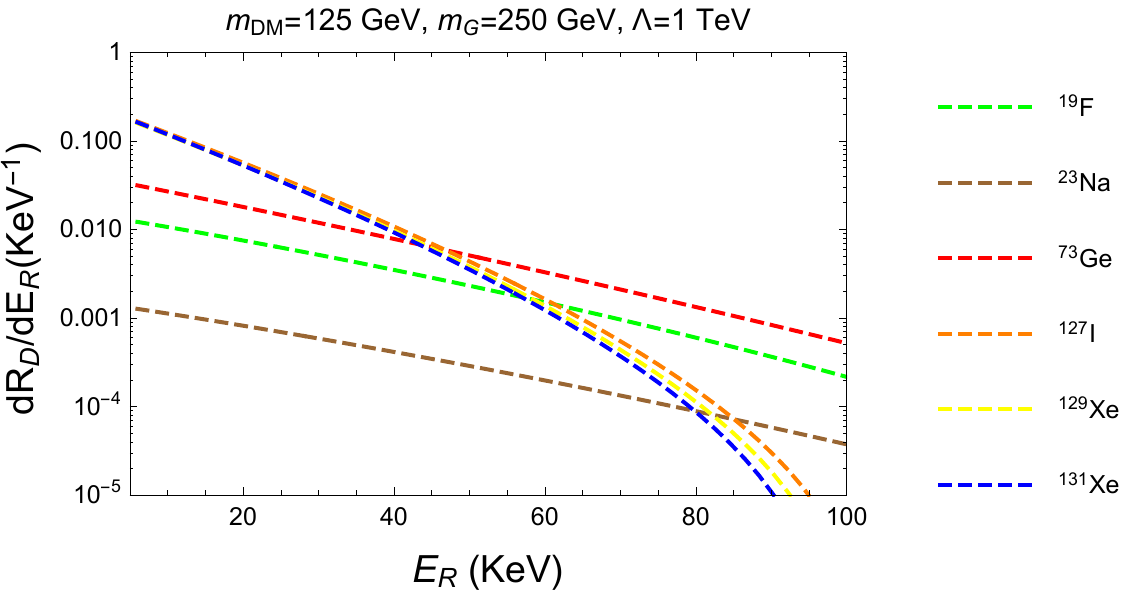}      
   \end{center}
       \caption{Differential event rates for fermionic dark matter (left) and scalar dark matter (right) for different experiments in Table \ref{mock} for $\Lambda=1$ TeV and $c_\chi =c_\psi =1$. }
  \label{FermionicEvents}
\end{figure}

\textsc{\begin{table}[t!]
\begin{center}
\begin{tabular}{|c|c|c|c|}
\hline
Nucleus   & Z & A & Exposure (Kg-day) \\ 
\hline 
F & 9 & 19 & 200000\\ 
\hline
Na & 11 & 23 & 14000\\ 
\hline
Ge & 32 & 73 & 36500\\ 
\hline
I & 53  & 127 & 78000\\ 
\hline
Xe & 54 & 129 & 73000\\ 
\hline
Xe & 54 & 131 & 73000\\ 
\hline
\end{tabular}
\caption{Mock experiments considered for the computation of differential scattering event rates in this model.}
\label{mock}
\end{center}
\end{table}}

In this section, we discuss the differential event rates for the spin-independent scattering between dark matter and nucleus in our model, for mock and current experiments of dark matter direct detection.

To compute the differential scattering event rates in our model, we take the model parameters that are consistent with the limits from DM direct detection experiments and use the package called DMFormFactor \cite{fitz,fitz2}.  A short review on the differential scattering event rates is given in Appendix A.

\begin{figure}[t!]
  \begin{center}
      \includegraphics[width=0.48\textwidth]{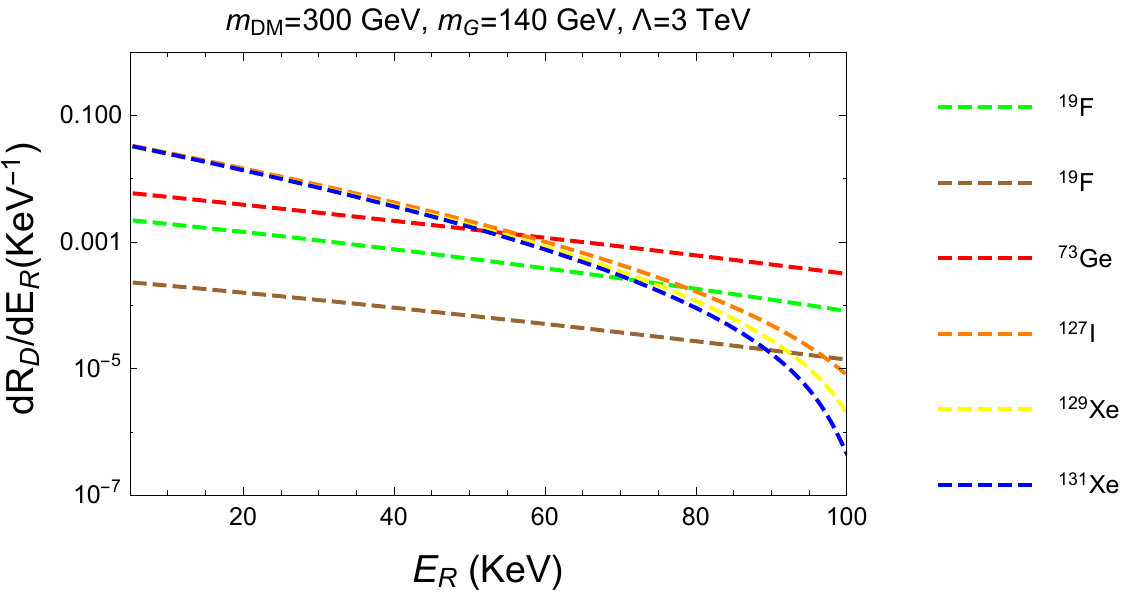} 
 		\includegraphics[width=0.48\textwidth]{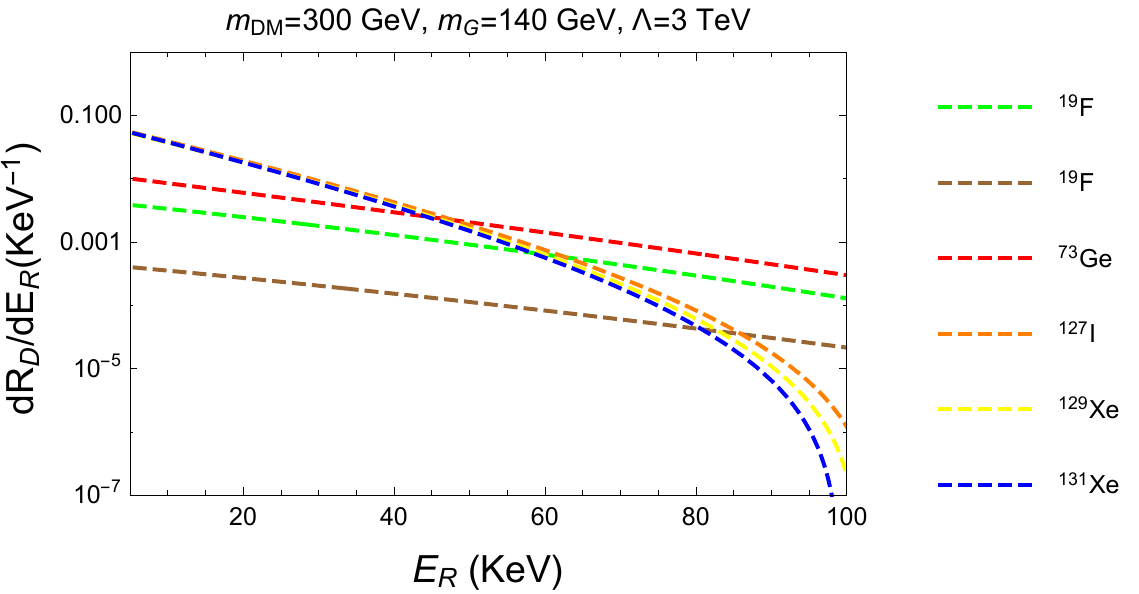}      
      \\
      \vspace{0.15cm}
          \      \includegraphics[width=0.48\textwidth]{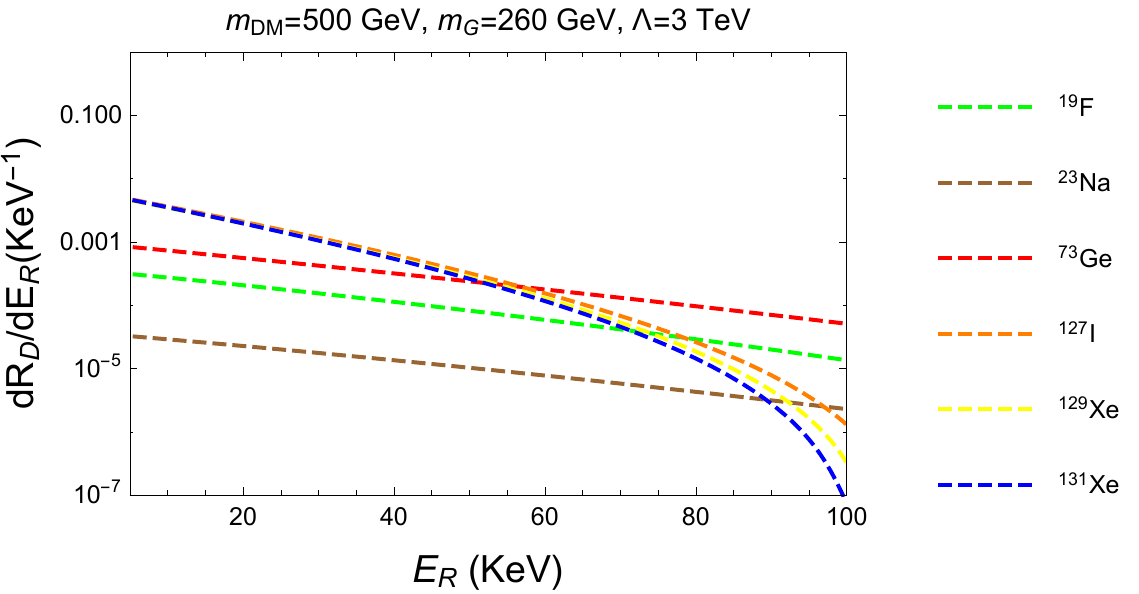} 
 		\includegraphics[width=0.48\textwidth]{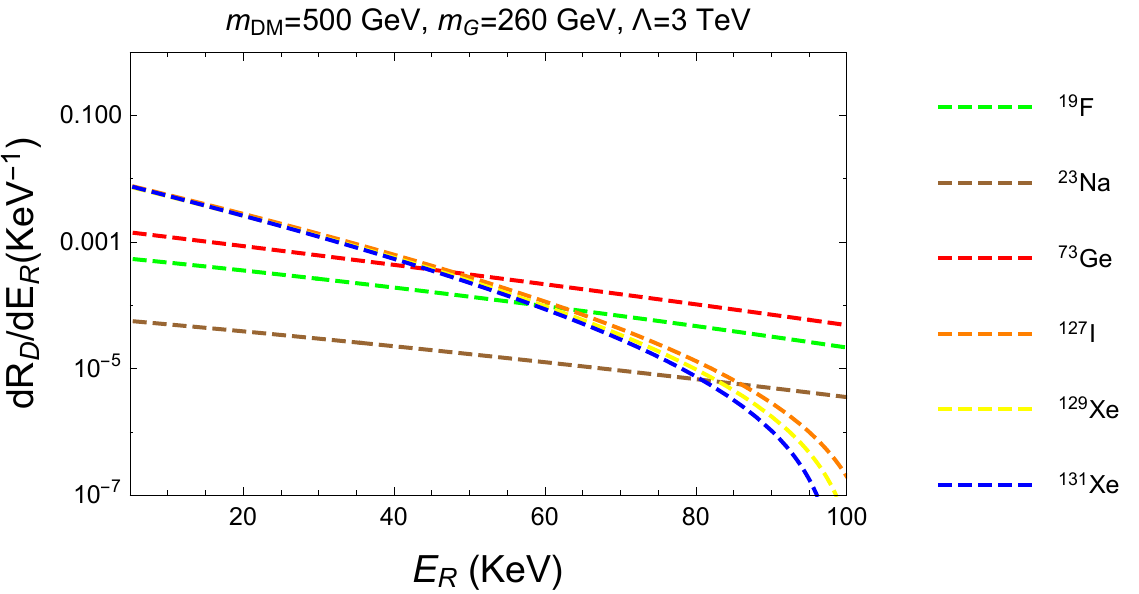}      
   \end{center}

  \caption{The same as in Fig.~\ref{FermionicEvents}, but with different masses for DM and spin--2 mediators and $\Lambda=3$ TeV.}
  \label{FermionicEvents3T}
\end{figure}

The input parameters for the package of DMFormFactor \cite{fitz,fitz2} are the spin and mass of DM, the information about the Galactic Halo (such as the escape velocity and the local DM density), our model parameters such as the couplings and mass of the graviton and the scale $\Lambda$, and finally the information about the detector we are considering. In our case, we use the parameters for different mock experiments with some of the most relevant isotopes as shown in Table~\ref{mock}.
Using the information in Table~\ref{mock}, the Lagrangians for the interactions in (\ref{feff2}) and (\ref{seff2}) and taking a zero momentum transfer $q\rightarrow 0$ approximation, we obtain the results for the differential event rates as a function of the recoil energy ($E_R$) in units of $\rm keV$ as in Figs.~(\ref{FermionicEvents}) and (\ref{FermionicEvents3T}), for the cases with fermionic and scalar dark matter for $\Lambda=1\,{\rm TeV}$ and $3\,{\rm TeV}$, respectively. For the fermionic case, the last operator in the Lagrangian ${\mathcal{O}}_3^{\rm NR} {\mathcal{O}}_5^{\rm NR}$ is a new type of interaction term that is allowed when the mediator is a spin 2 particle. But, the ${\mathcal{O}}_3^{\rm NR} {\mathcal{O}}_5^{\rm NR}$  term is velocity-suppressed so it is not included in our study. Therefore, the differential event rates for fermion and scalar dark matter are similar when the DM mass and the mass and coupling of the spin-2 mediator are the same.
However, as will be shown in the next section, the annihilation cross sections of dark matter crucially depend on the spin of dark matter. 

\begin{figure}[t!]
  \begin{center}
      \includegraphics[width=0.48\textwidth]{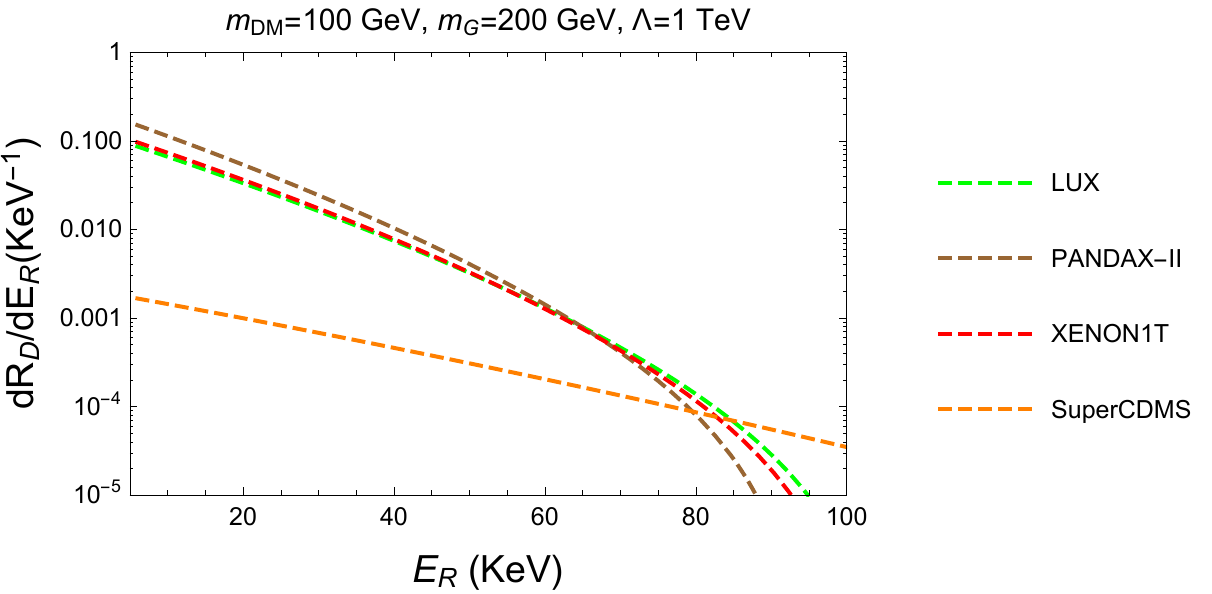} 
 		\includegraphics[width=0.48\textwidth]{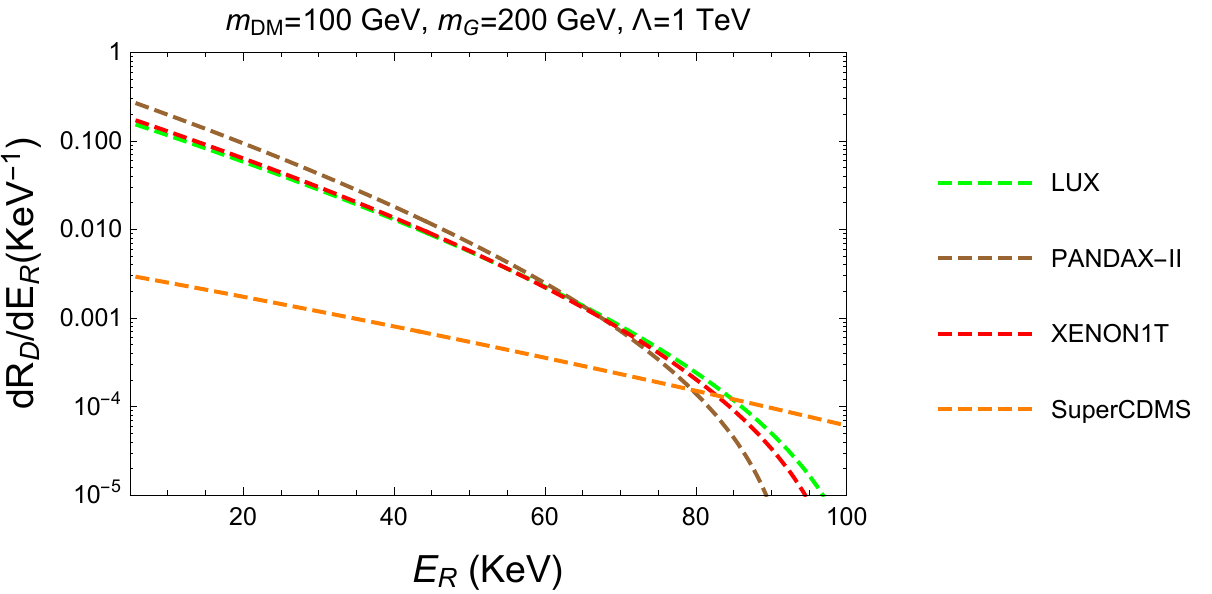}      
      \\
      \vspace{0.15cm}
          \      \includegraphics[width=0.48\textwidth]{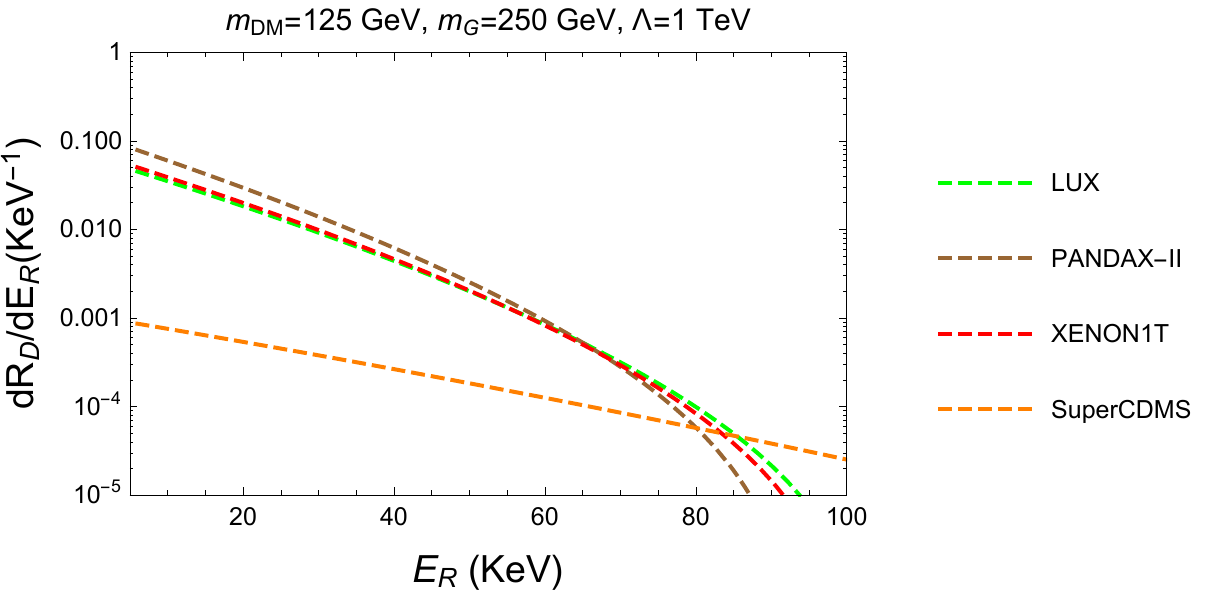} 
 		\includegraphics[width=0.48\textwidth]{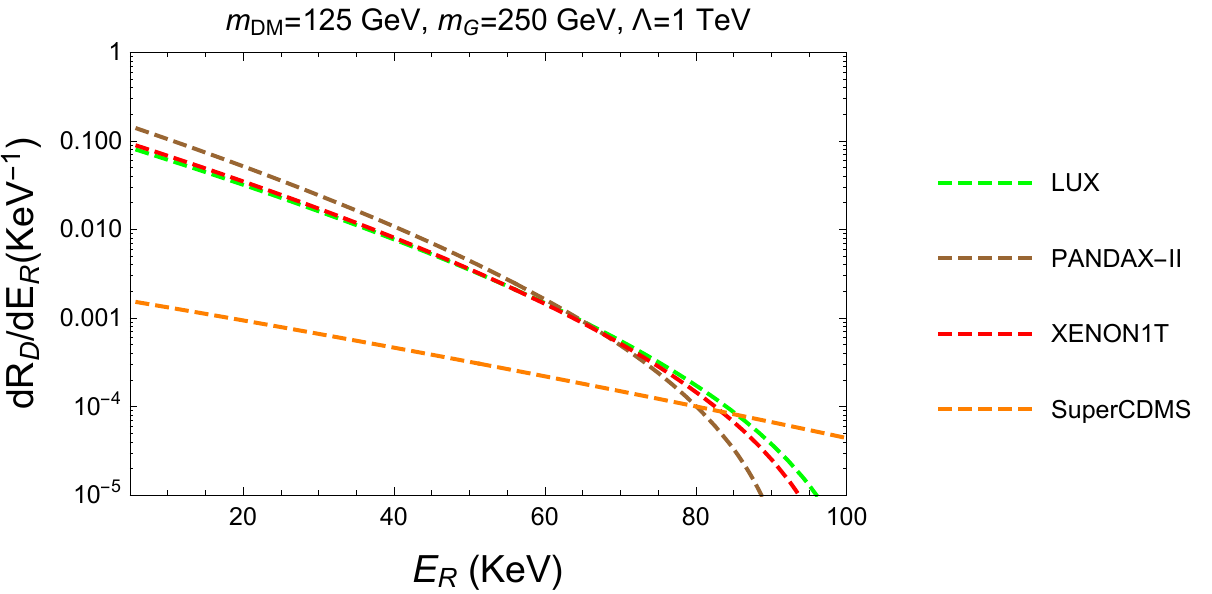}       
   \end{center}
       \caption{Differential event rates for fermionic dark matter (left) and scalar dark matter (right) for current experiments for $\Lambda=1$ TeV and $c_\chi =c_\psi =1$. }
  \label{FermionicEventsE}
\end{figure}

\begin{figure}[t!]
  \begin{center}
      \includegraphics[width=0.48\textwidth]{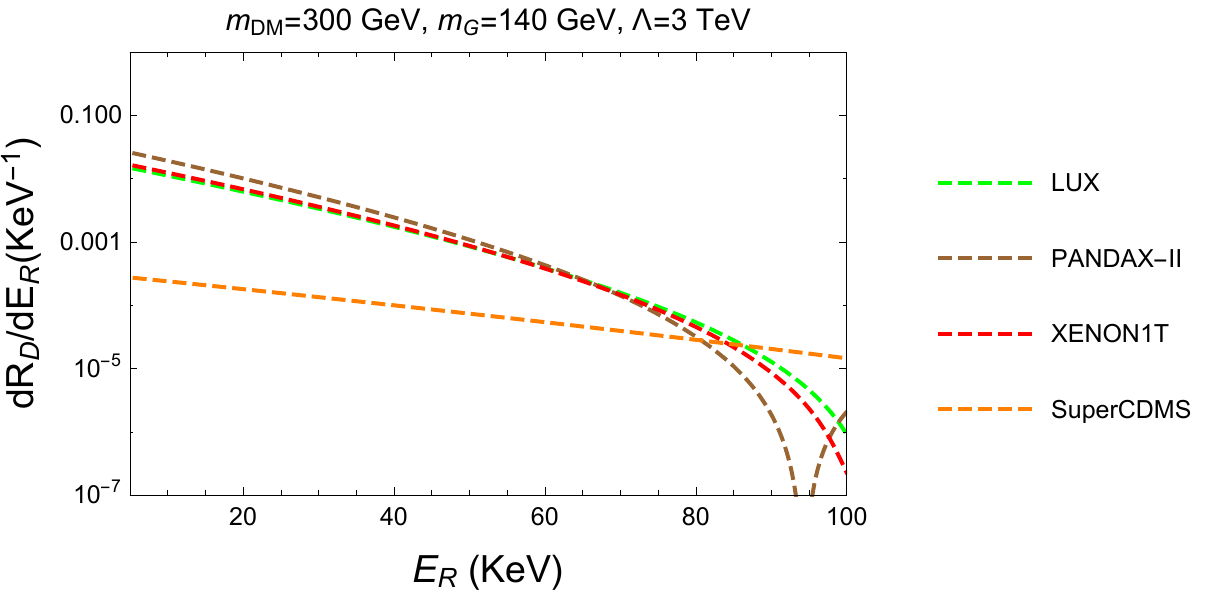} 
 		\includegraphics[width=0.48\textwidth]{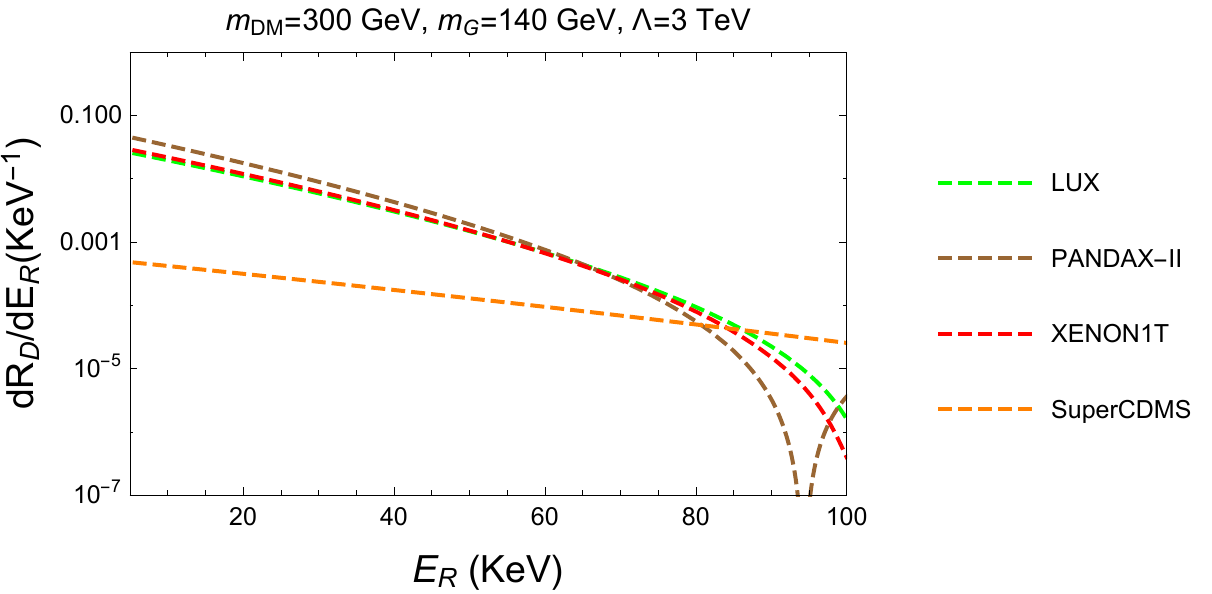}      
      \\
      \vspace{0.15cm}
          \      \includegraphics[width=0.48\textwidth]{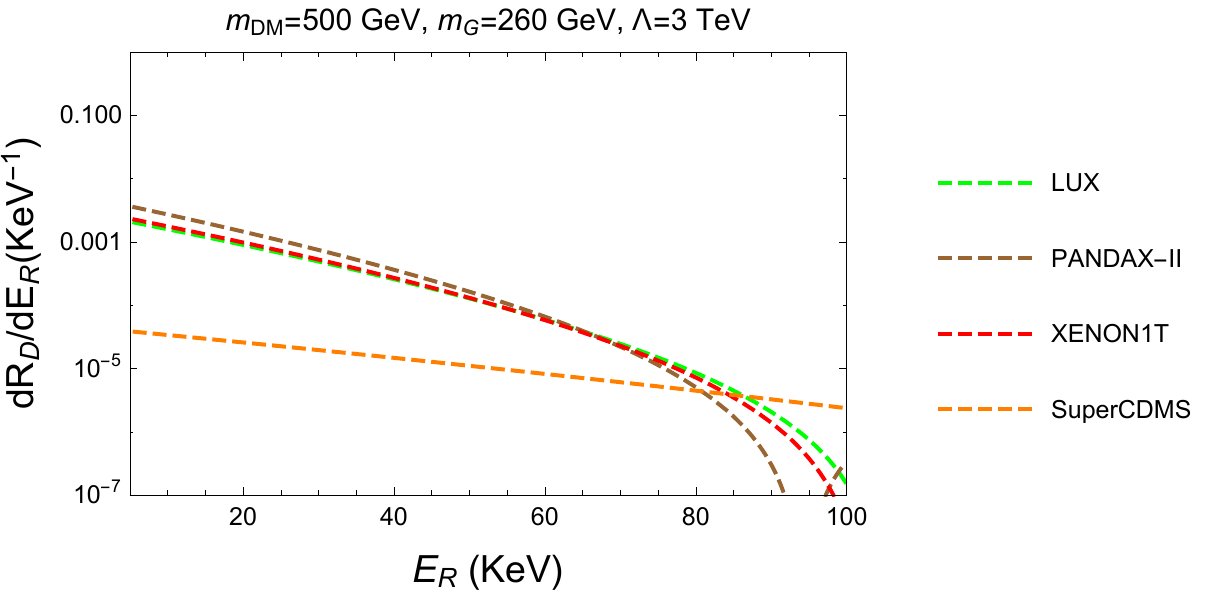} 
 		\includegraphics[width=0.48\textwidth]{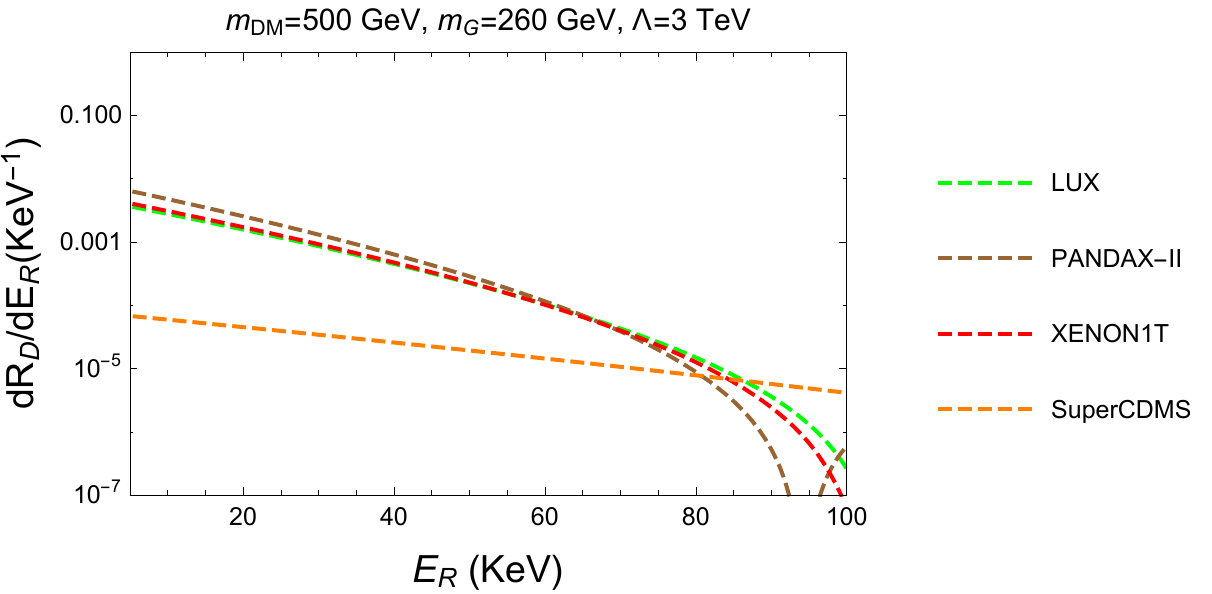}      
   \end{center}

  \caption{The same as in Fig.~\ref{FermionicEventsE}, but with different masses for DM and spin--2 mediators and $\Lambda=3$ TeV. }
  \label{FermionicEvents3TE}
\end{figure}

\textsc{\begin{table}[t!]
\begin{center}
\begin{tabular}{|c|c|c|c|}
\hline
Experiment (Nucleus)  & Z & A & Exposure (Kg-day) \\ 
\hline 
LUX (Xe) & 54 & 129 & 33500\\ 
\hline
XENON1T (Xe)& 54 & 131 & 36500\\ 
\hline
PandaX-II (Xe)& 54 & 136 & 54000\\ 
\hline
SuperCDMS (Ge)& 32  & 73 & 1690\\ 
\hline
CDMSlite (Ge)& 32  & 73 & 70\\ 
\hline
XENON10 (Xe)& 54  & 131 & 15\\ 
\hline
DarkSide-50 (Ar)& 18  & 39 & 46\\ 
\hline
\end{tabular}
\caption{Detector information for the current experiments considering in this study for the computation of differential scattering event rates in this model.}
\label{mock2}
\end{center}
\end{table}}

Also, we obtained similar plots, considering the detectors used in current DM experiments as XENON1T \cite{xenon1t}, PandaX-II \cite{panda}, SuperCDMS \cite{cdms}, LUX \cite{lux}, CDMSlite \cite{CDMSlite}, XENON10 \cite{xenon10}, and DarkSide-50 \cite{darkside50}, with the detector parameters shown in Table~\ref{mock2}. Some results for differential event rates with WIMP dark matter are shown in Figs.~\ref{FermionicEventsE} and \ref{FermionicEvents3TE}, for $\Lambda=1\,{\rm TeV}$ and $3\,{\rm TeV}$, respectively, and with the parameters that are consistent with relic density condition, ATLAS dijet and direct detection bounds, as will be discussed in the next section.

\section{Bounds from relic density and direct detection}

We consider the annihilation cross sections for fermion, scalar or vector dark matter in order to determine the relic density. Then, we discuss the relic density condition for the parameter space of our model and also impose the direct detection  limits on the total spin-independent elastic scattering cross section and the dijet bounds on the spin-2 mediator.

\subsection{Fermion dark matter}

The annihilation cross section for $\chi{\bar\chi}\rightarrow \psi{\bar\psi}$ is given \cite{GMDM1,GMDM2,diphoton} by 
\bea
(\sigma v)_{\chi{\bar\chi}\rightarrow \psi{\bar\psi}} = v^2 \cdot \frac{N_c c^2_\chi c^2_\psi }{72\pi\Lambda^4}
\frac{m^6_\chi}{(4m^2_\chi-m^2_G)^2+\Gamma^2_G m^2_G} \left(1-\frac{m^2_\psi}{m^2_\chi}\right)^\frac{3}{2} 
\left(3+\frac{2m^2_\psi}{m^2_\chi}\right).
\eea
Thus, the annihilation of fermion dark matter into quarks becomes $p$-wave suppressed. 
When $m_{\chi}>m_G$, there is an extra contribution to the annihilation cross section, due to the $t$-channel for both models \cite{GMDM1,GMDM2,diphoton}, as follows,
\bea
(\sigma v)_{\chi \bar\chi \rightarrow GG} &=& \frac{c_{\chi}^4 m_{\chi}^2}{16 \pi \Lambda^4 }
\frac{(1-r_\chi)^\frac{7}{2}}{r^4_\chi (2-r_\chi)^2}  \label{tch-fermion}
\eea
with $r_\chi = \left(\frac{m_G}{m_\chi}\right)^2$. Then,   the $t$-channel annihilation is $s$-wave, so it becomes dominant in determining the relic density for heavy fermion dark matter.

\subsection{Scalar dark matter}

The annihilation cross section for $SS\rightarrow \psi{\bar\psi}$ is given \cite{GMDM1,GMDM2,diphoton} by 
\bea
(\sigma v)_{SS\rightarrow\psi{\bar\psi} } = v^4 \cdot  \frac{ N_c c_S^2 c_\psi^2 }{360\pi \Lambda^4} 
\frac{m_S^6}{(m_G^2-4 m_S^2)^2+\Gamma_G^2 m_G^2}
\left(1-\frac{m_\psi^2}{m_S^2}\right)^\frac{3}{2} \left(3+\frac{2m_\psi^2}{m_S^2}\right) .
\eea
Thus, the annihilation of scalar dark matter into quarks becomes $d$-wave suppressed. 

When $m_S>m_G$, there is an extra contribution to the annihilation cross section, due to the $t$-channel for both models \cite{GMDM1,GMDM2,diphoton}, as follows,
\bea
(\sigma v)_{SS\rightarrow GG} = \frac{4 c_{S}^4 m_{S}^2}{9 \pi \Lambda^4 }
\frac{(1-r_S)^\frac{9}{2}}{r^4_S  (2-r_S)^2}   \label{tch-scalar}
\eea
with $r_S = \left(\frac{m_G}{m_S}\right)^2$.

\subsection{Vector dark matter}

The annihilation cross section for $XX\rightarrow \psi{\bar\psi}$ is given \cite{GMDM1,GMDM2,diphoton} by
\bea
(\sigma v)_{XX\rightarrow \psi{\bar\psi}}&=& \frac{4N_c c^2_X c^2_\psi }{27\pi \Lambda^4}
 \frac{m^6_X}{(4m^2_X-m^2_G)^2+\Gamma^2_G m^2_G}\left(3+\frac{2m^2_\psi}{m^2_X}\right)\left(1-\frac{m^2_\psi}{m^2_X}\right)^\frac{3}{2}.
\eea
Thus, the annihilation of vector dark matter into quarks becomes $s$-wave suppressed. 
In this case, smaller spin-2 mediator couplings to the SM quarks or vector dark matter can be consistent with the correct relic density, as compared to the other cases. 
But, indirect detection signals from the annihilation of vector dark matter are promising \cite{GMDM2}.

For $m_X>m_G$, there is an extra contribution to the annihilation cross section, due to the $t$-channel in both models \cite{GMDM1,GMDM2,diphoton}, as follows,
\bea
(\sigma v)_{X X  \rightarrow GG} &=&
\frac{c_{X}^4 m_{X}^2}{324 \pi \Lambda^4 }
\frac{\sqrt{1-r_X}}{r^4_X  (2-r_X)^2} \,
\bigg(176+192 r_X+1404 r^2_X-3108 r^3_X \nonumber \\
&&+1105 r^4_X+362 r^5_X+34 r^6_X \bigg)
\eea
with $r_X = \left(\frac{m_G}{m_X}\right)^2$.

\subsection{Bounds on WIMP dark matter}

Taking a zero momentum transfer for the DM-nucleon scattering, we 
use the nucleon matrix elements for twist-2 operators given in eq.~(\ref{twist2-0}) or the results in Appendix C  and simply obtain the total cross section for spin-independent elastic scattering between dark matter and nucleus as
\bea
\sigma_{{\rm DM}-A}^{ SI}= \frac{ \mu^2_A}{\pi}\,\Big(Z f^{\rm DM}_p+(A-Z) f^{\rm DM}_n\Big)^2
\eea
where $\mu_A=m_\chi m_A/(m_\chi +m_A)$ is the reduced mass of the DM-nucleus system and
$m_A$ is the target nucleus mass, $Z,A$ are the number of protons and the atomic number, respectively, and the nucleon form factors are given by the same formula for all the spins of dark matter as
\bea
f^{\rm DM}_p&=& \frac{ c_{\rm DM} m_N m_{\rm DM}}{4m^2_G\Lambda^2}\bigg(\sum_{\psi=u,d,s,c,b}3c_\psi(\psi(2)+{\bar\psi}(2))+\sum_{\psi=u,d,s} \frac{1}{3} c_\psi f^p_{T\psi}\bigg), \\
f^{\rm DM}_n&=& \frac{c_{\rm DM}  m_N m_{\rm DM}}{4m^2_G\Lambda^2}\bigg(\sum_{\psi=u,d,s,c,b}3c_\psi(\psi(2)+{\bar\psi}(2))+\sum_{\psi=u,d,s} \frac{1}{3} c_\psi f^n_{T\psi}\bigg)
\eea
where ${\rm DM}=\chi, S, X$  for fermion, scalar and vector dark matter, respectively.
The results are the same as those for the general effective interactions with momentum transfer taken to zero, in eq.~(\ref{feff2}), (\ref{seff2}) and (\ref{veff}).

\begin{figure}[t!]
  \begin{center}
      \includegraphics[height=0.31\textwidth]{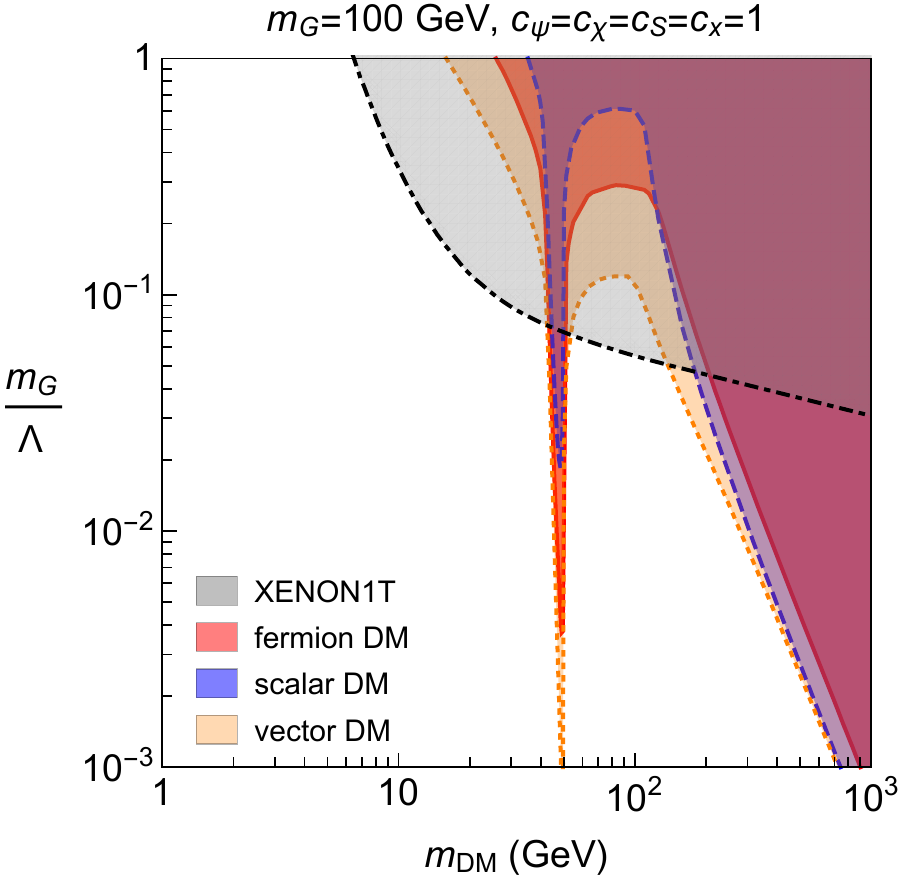}
      \includegraphics[height=0.31\textwidth]{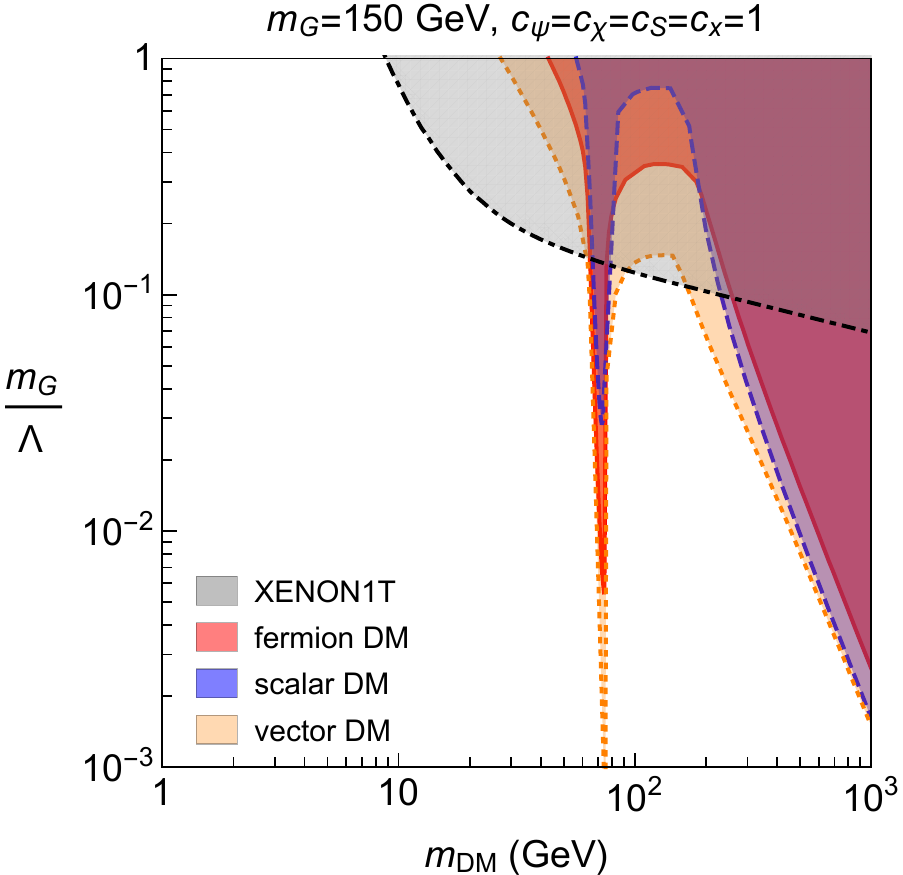}
       \includegraphics[height=0.31\textwidth]{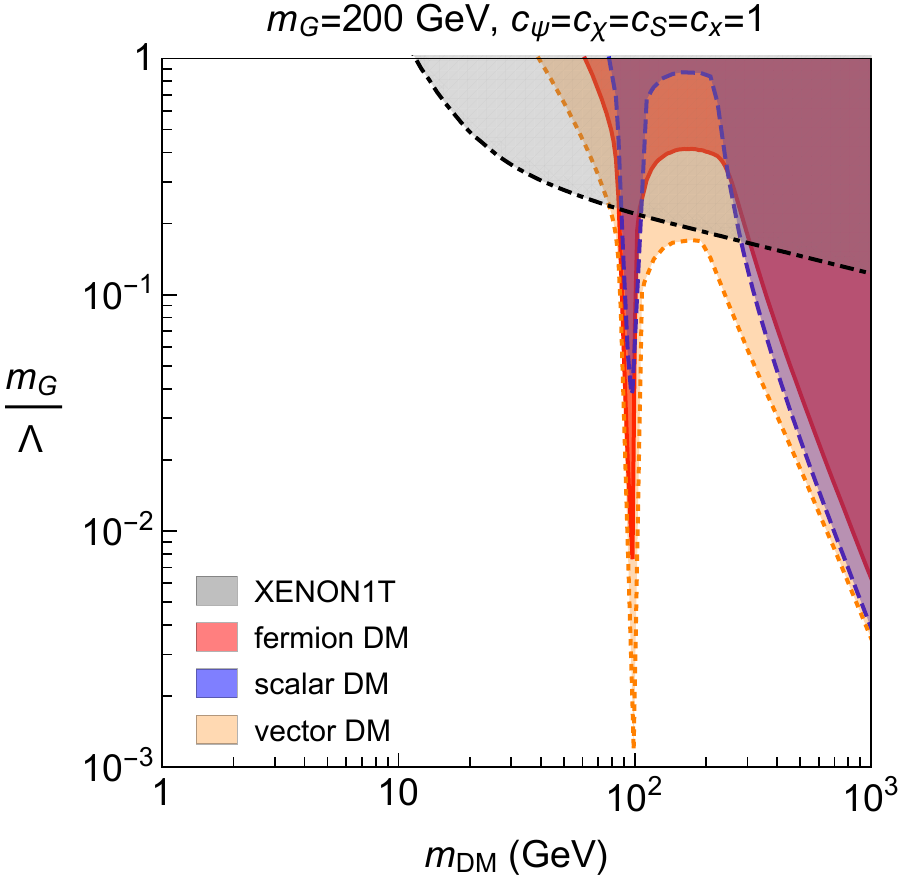}
   \end{center}
  \caption{Parameter space of fermion and scalar dark matter in $m_{\rm DM}$ vs $m_G/\Lambda$. The gray regions are excluded by XENON1T. We took $c_\chi=c_S=c_{u,d,s,c,b,t}=1$ and $m_G=100, 150, 200\,{\rm GeV}$ on left, middle and right, respectively.  }
  \label{relic1}
\end{figure}

The most relevant isotopes in direct detection experiments are $^{129,131}{\rm Xe}$, $^{127}{\rm I}$, $^{73}{\rm Ge}$, $^{19}{\rm F}$, $^{23}{\rm Na}$, $^{27}{\rm Al}$ and $^{29}{\rm Si}$.
For instance, we get $Z=54$ and $A-Z=75$ for $^{129}{\rm Xe}$.
The above DM-nucleus scattering cross section is related to the normalized-to-proton scattering cross section  $\sigma^{SI}_{{\rm DM}-p}$, that is usually presented for experimental limits, by  $\sigma^{SI}_{{\rm DM}-p}=(\mu_N/\mu_A)^2\sigma^{\rm SI}_{{\rm DM}-A}/A^2$ with $\mu_N=m_{\rm DM} m_N/(m_{\rm DM}+m_N)$.

In Fig~\ref{relic1}, we depict in the parameter space for $m_{\rm DM}$ vs $m_G/\Lambda$  the region where the DM relic density overcloses the Universe in red, blue and orange for fermion, scalar and vector dark matter, respectively. The regions in gray are ruled out by the direct detection experiment in XENON1T \cite{xenon1t}. We have taken $m_G=100, 150, 200\,{\rm GeV}$  from left to right plots and the couplings of DM and quarks to the spin-2 mediator are the same as $c_\chi=c_S=c_{u,d,s,c,b,t}=1$ in all the plots.
As a result, we find that the non-resonance regions saturating the relic density, away from the resonance with $m_G\sim 2m_{\rm DM}$, are tightly constrained by XENON1T bounds.
The non-resonance regions below $m_{\rm DM}=200-300\,{\rm GeV}$ have been already excluded but the non-resonance regions with larger DM masses and the resonance region are less constrained by the current data. In particular, for $m_{\rm DM}>m_G$, dark matter can annihilate into a pair of spin-2 mediators so there is no need of a sizable coupling between dark matter and SM fermions for the correct relic density. Therefore, those regions can be probed by updated XENON1T and future direct detection experiments.

\begin{figure}[t!]
  \begin{center}
      \includegraphics[height=0.32\textwidth]{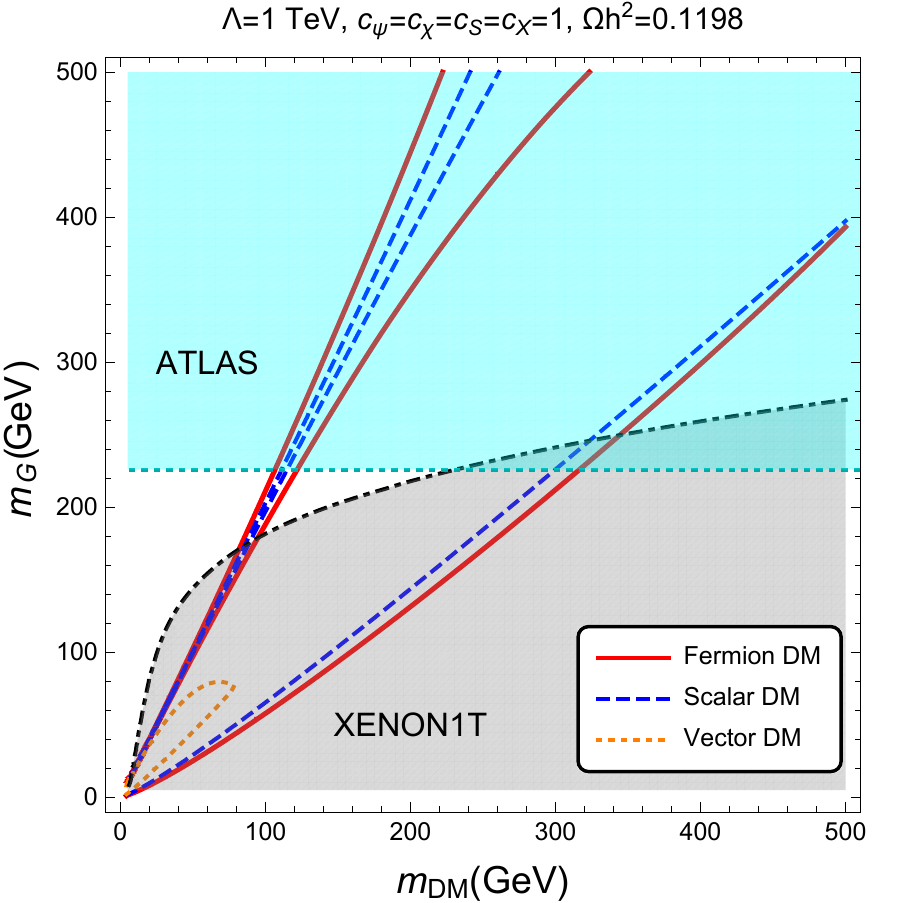}
      \includegraphics[height=0.32\textwidth]{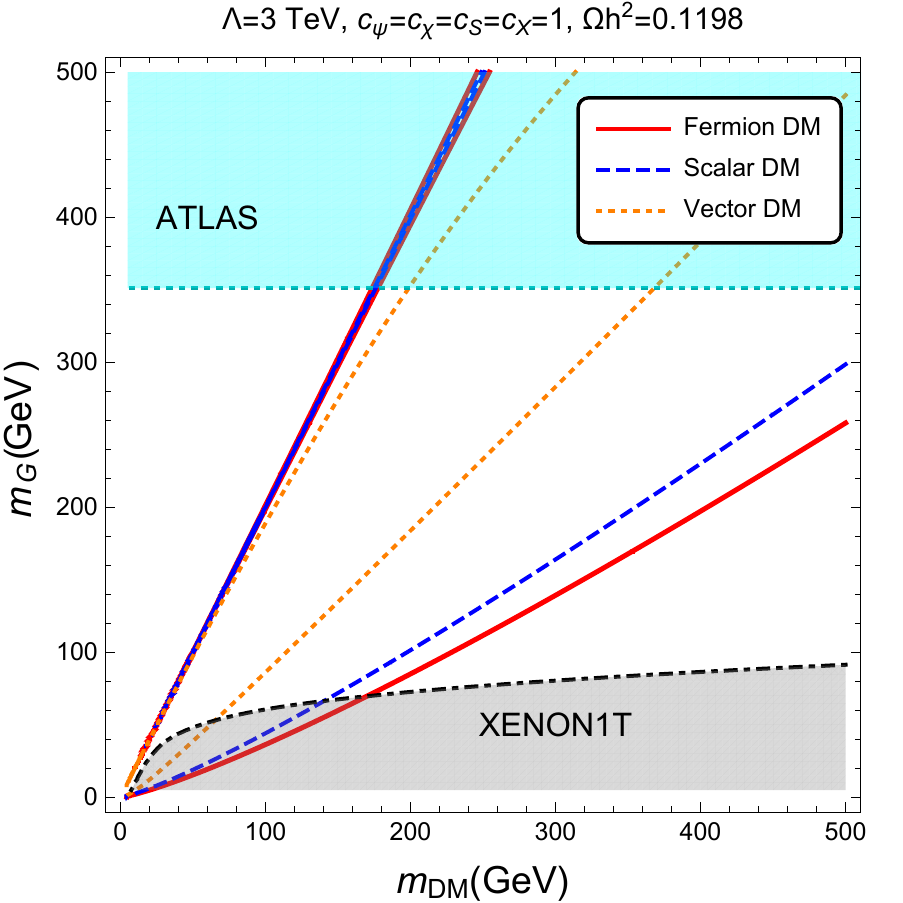}
       \includegraphics[height=0.32\textwidth]{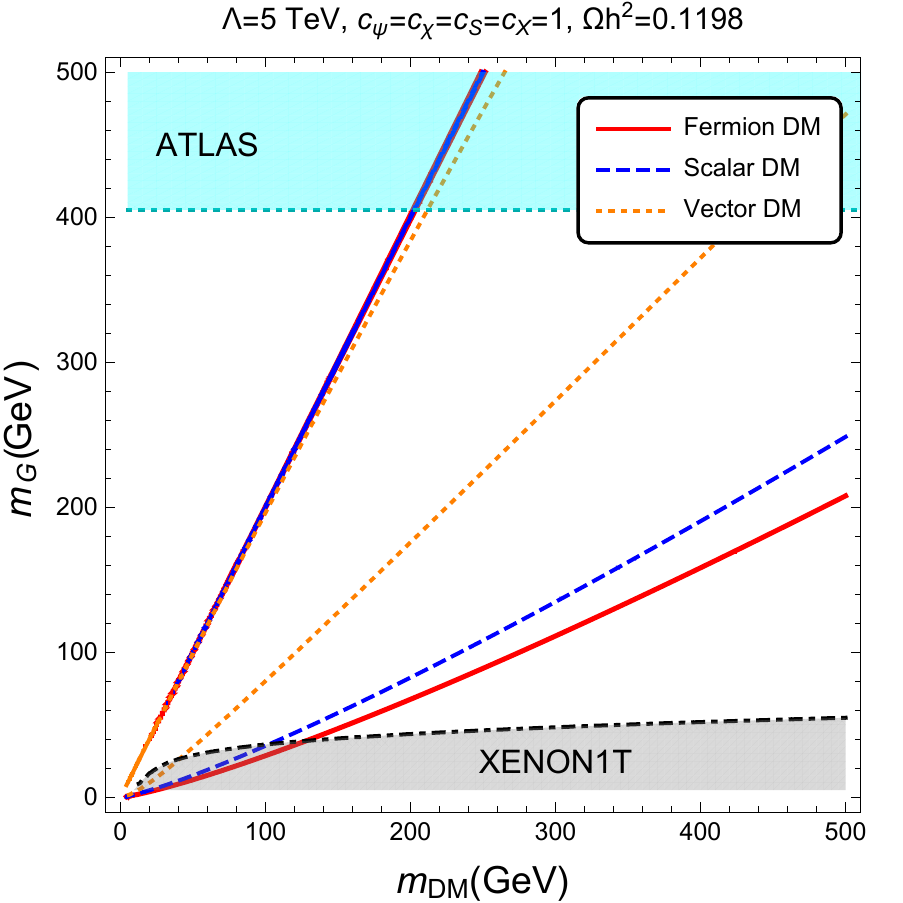}
   \end{center}
  \caption{Parameter space of fermion and scalar dark matter in $m_{\rm DM}$ vs $m_G$.  The gray and cyan regions are excluded by XENON1T and ATLAS dijet searches, respectively. We took $\Lambda=1,3,5\,{\rm TeV}$ on left, middle and right, respectively. The other parameters are the same as in Fig.~\ref{relic1}. }
  \label{relic2}
\end{figure}

In Fig.~\ref{relic2}, we impose in the parameter space for $m_{\rm DM}$ vs $m_G$ the same conditions from the relic density and the limits from XENON1T. The relic density is saturated by the DM annihilation into quarks along the red, blue and orange lines, for fermion, scalar and vector dark matter, respectively. The regions in gray are ruled out by the direct detection experiment in XENON1T \cite{xenon1t}. 
We also overlaid in cyan regions the bounds from dijet resonance searches with mono-photon at the LHC \cite{atlas}.  In the case with $m_G>2m_{\rm DM}$, for which the spin-2 mediator decays invisibly into a pair of dark matter, the ATLAS dijet limit on $\Lambda$ scales by $\sqrt{{\rm BR}(G\rightarrow q{\bar q})}=\sqrt{\frac{15}{19}}\Big(\sqrt{\frac{15}{16}}\Big)$ with $q=u,d,s,c,b$ for $m_G>2m_t$($m_G<2m_t$), which leads only to a very mild change in the cyan region in Fig.~\ref{relic2}. We have taken $\Lambda=1, 3, 5\,{\rm TeV}$  from left to right plots and the same couplings of DM and quarks to the spin-2 mediator as $c_\chi=c_S=c_{u,d,s,c,b,t}=1$ in all the plots.
In the case with $\Lambda=1\,{\rm TeV}$, the WIMP parameter space, in particular, the non-resonance region, is tightly constrained by both XENON1T and dijet bounds. But,  for larger values of $\Lambda=3, 5\,{\rm TeV}$, a wide parameter space opens up and can be tested by updated XENON1T and future experiments.

\subsection{Bounds on light dark matter}

\begin{figure}[t!]
  \begin{center}
      \includegraphics[width=0.48\textwidth]{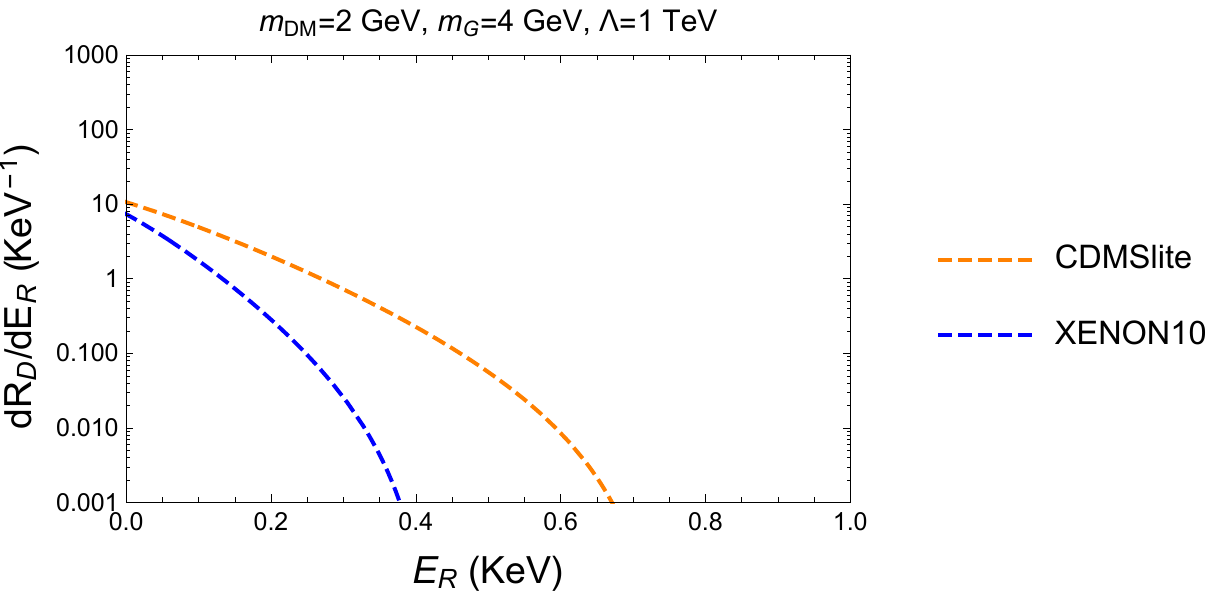} 
 		\includegraphics[width=0.48\textwidth]{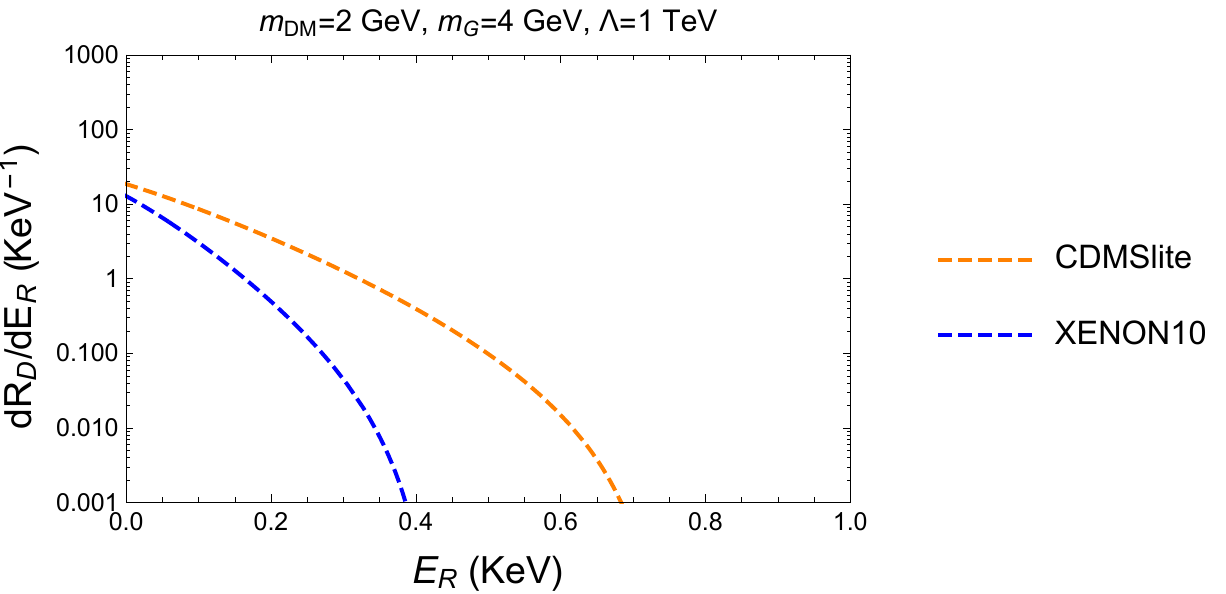}      
      \\       \vspace{0.15cm}
      \includegraphics[width=0.48\textwidth]{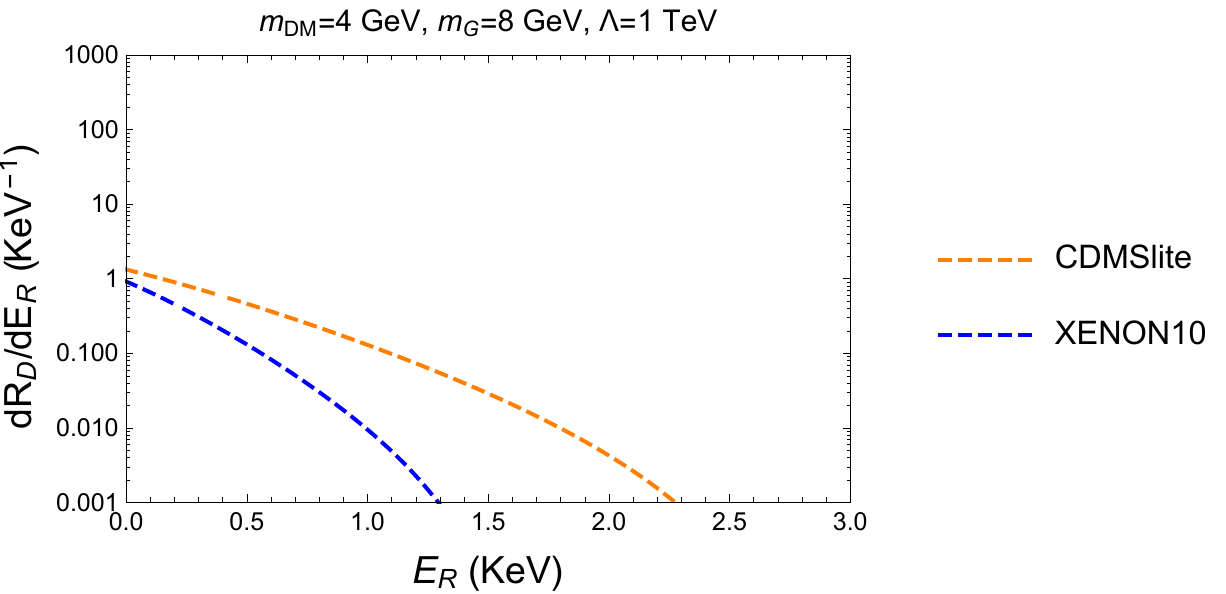} 
 		\includegraphics[width=0.48\textwidth]{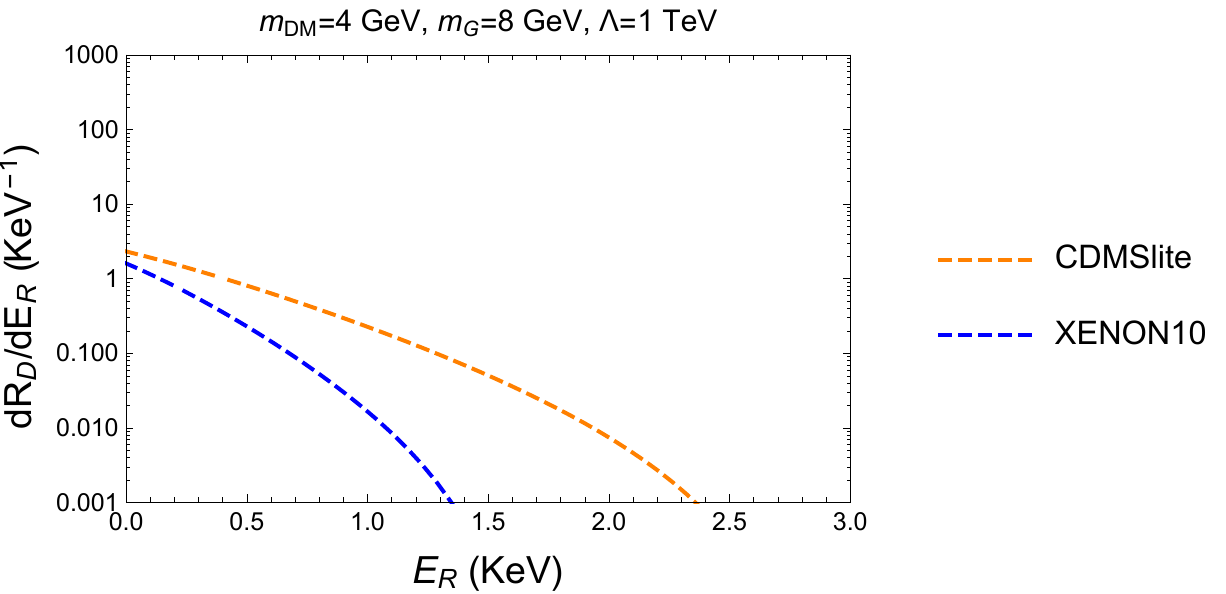}      
   \end{center}

  \caption{Differential event rates for light fermion (left) or scalar (right) dark matter for current experiments for $\Lambda=1$ TeV and $c_\chi =c_\psi =1$. }
  \label{lightDM-events}
\end{figure}

\begin{figure}[t!]
  \begin{center}
      \includegraphics[width=0.48\textwidth]{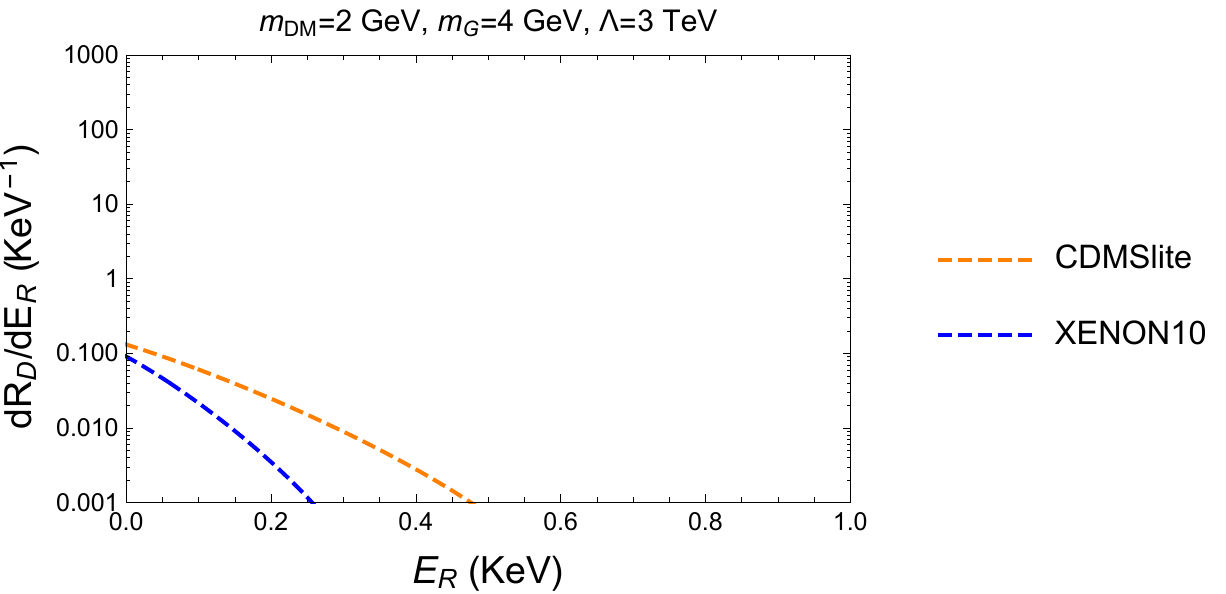} 
 		\includegraphics[width=0.48\textwidth]{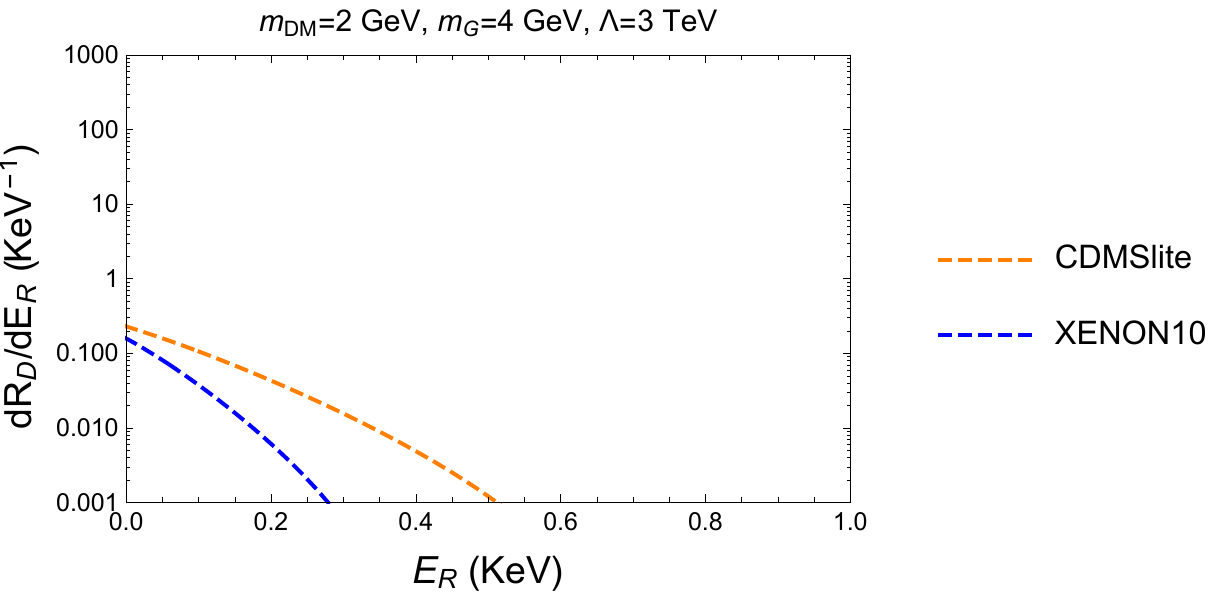}      
      \\       \vspace{0.15cm}
      \includegraphics[width=0.48\textwidth]{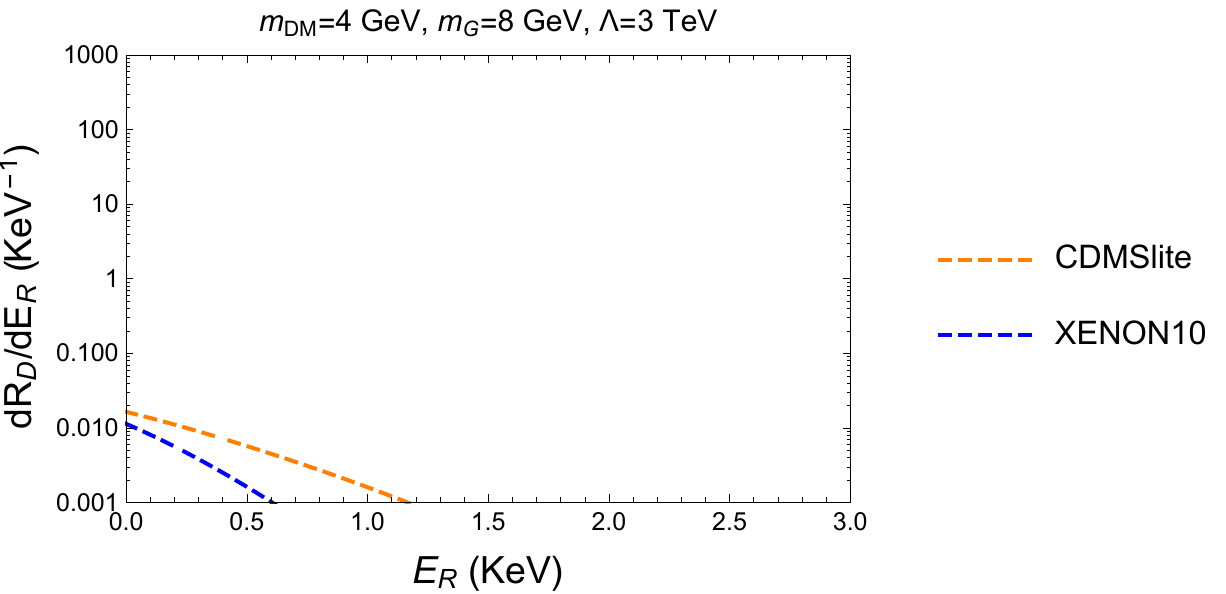} 
 		\includegraphics[width=0.48\textwidth]{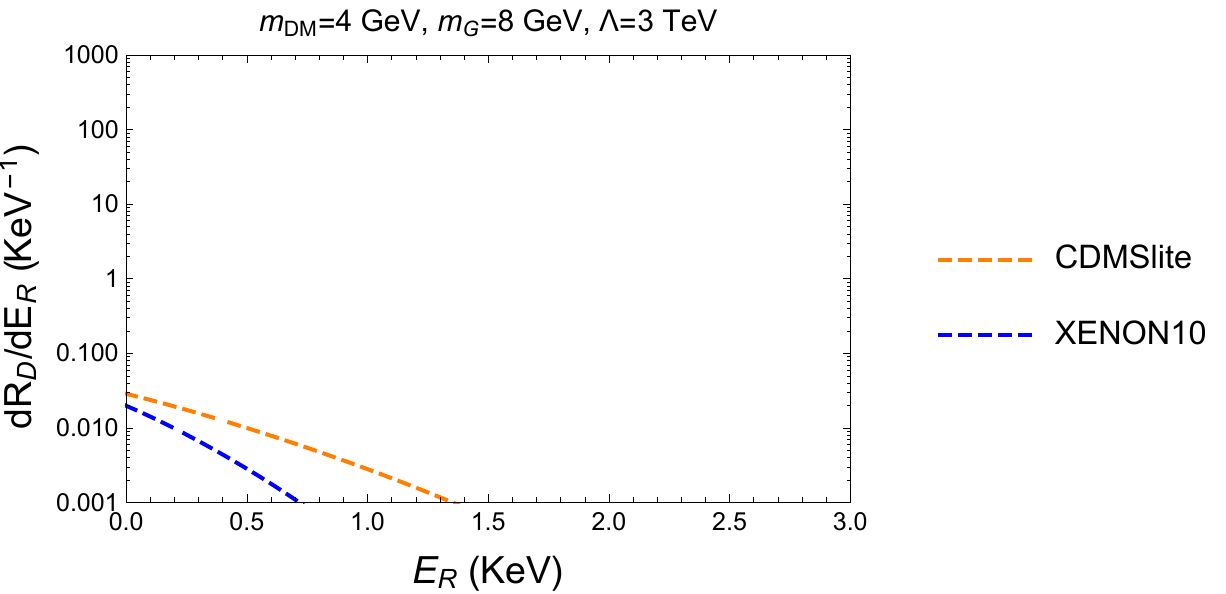}      
   \end{center}

  \caption{The same as in Fig.~\ref{lightDM-events}, but for $\Lambda=3$ TeV. }
  \label{lightDM-events2}
\end{figure}

Some results for the corresponding differential event rates with light fermion or scalar dark matter below $10\,{\rm GeV}$ are shown for CDMSlite and XENON10 experiments in  Figs.~\ref{lightDM-events} and \ref{lightDM-events2}. Here, we have chosen the parameters that are consistent with direct detection bounds, in particular, from XENON10 and cryogenic direct detection experiments such as CDMSlite and CRESST.

\begin{figure}[t!]
  \begin{center}
      \includegraphics[height=0.32\textwidth]{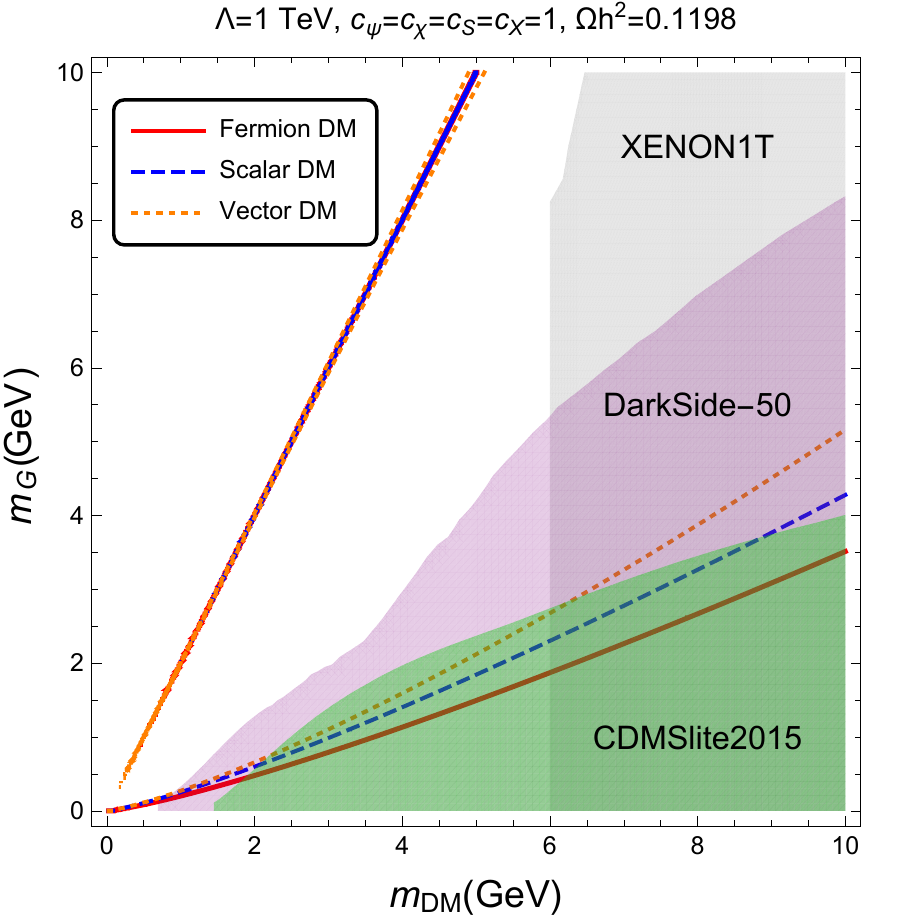}
      \includegraphics[height=0.32\textwidth]{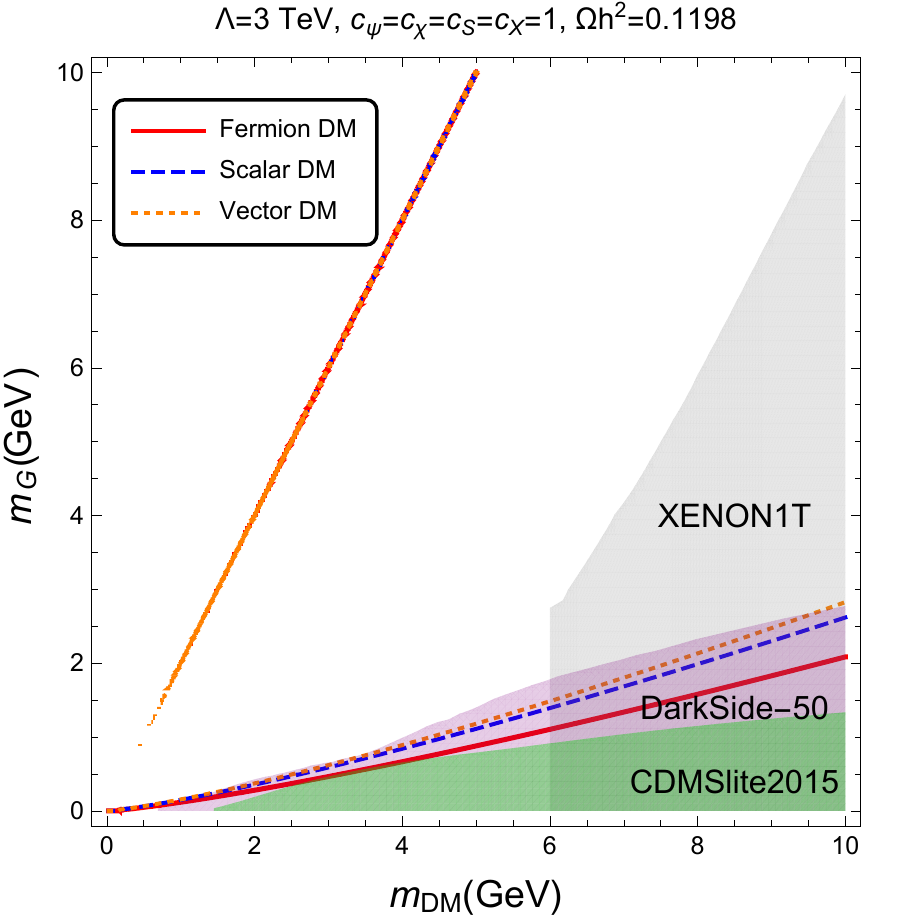}
       \includegraphics[height=0.32\textwidth]{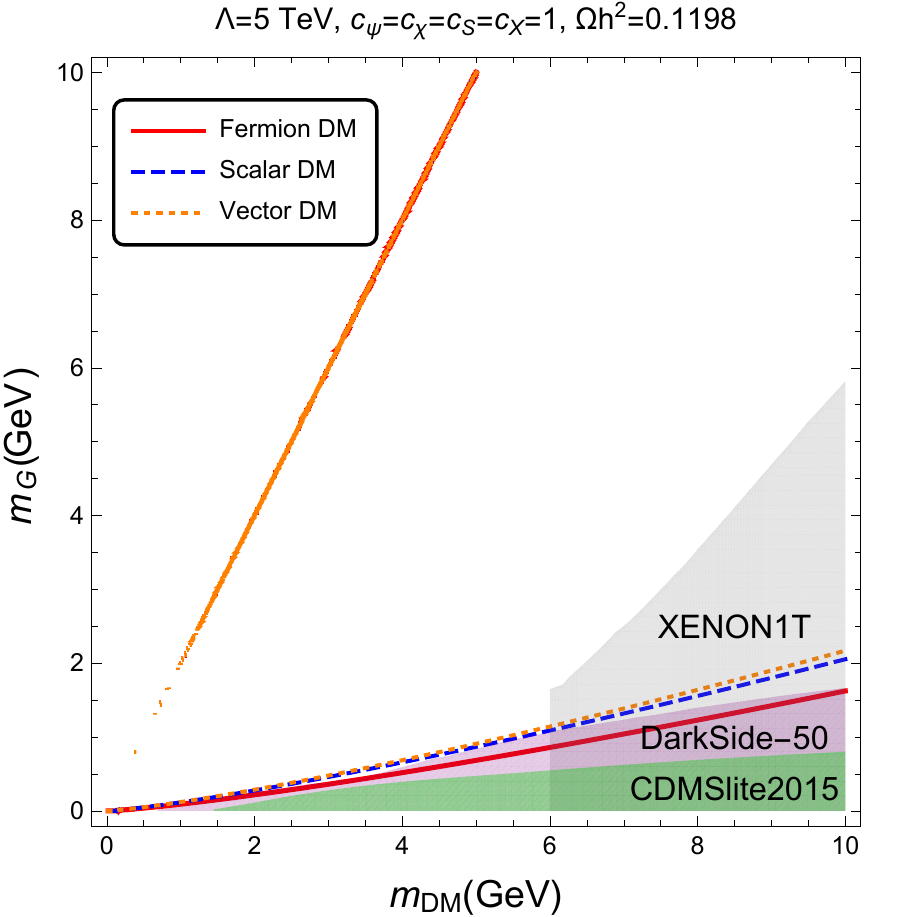}
   \end{center}
  \caption{Parameter space of light dark matter below $10\,{\rm GeV}$.  The gray, green and purple regions are excluded by XENON1T, CDMSlite and DarkSide-50, respectively.  We took $\Lambda=1,3,5\,{\rm TeV}$ on left, middle and right, respectively. The other parameters are the same as in Fig.~\ref{relic1}. }
  \label{relic3}
\end{figure}

In Fig.~\ref{relic3}, we considered the case with light dark matter of mass below $10\,{\rm GeV}$. In this case, cryogenic direct detection experiments \cite{lightDM} such as CDMSlite \cite{CDMSlite}, CRESST \cite{CRESST} and DarkSide-50 \cite{darkside50} with low thresholds for recoil energy are relevant for $m_{\rm DM}=1.45-9\,{\rm GeV}$, $0.71-9\,{\rm GeV}$, and $1.8-6\,{\rm GeV}$, respectively. We showed that XENON1T, CDMSlite and DarkSide-50 experiments rule out the parameter space in gray, green and purple regions, respectively. We note that the bounds from CRESST or XENON10 are less stringent that those from other experiments, so we don't show them in Fig.~\ref{relic3}.
We have taken $\Lambda=1, 3, 5\,{\rm TeV}$  from left to right plots and the couplings of DM and quarks to the spin-2 mediator are the same as $c_\chi=c_S=c_{u,d,s,c,b,t}=1$ in all the plots. 
 As a consequence, for a low $\Lambda=1\,{\rm TeV}$, the region where dark matter annihilation into a pair of spin-2 mediators explains the correct relic density is almost excluded by direct detection, except for $m_{\rm DM}\lesssim 2\,{\rm GeV}$. The  resonance region with $m_G\sim 2m_{\rm DM}$ survives the direct detection bounds.
On the other hand, for larger values of $\Lambda=3, 5\,{\rm TeV}$, the more non-resonance region below $m_{\rm DM}\simeq 6\,{\rm GeV}$ survives.

\section{Conclusions}

We have presented the effective interactions between dark matter and the SM quarks due to the massive spin-2 mediator. The resulting non-relativistic operators for nucleons have specific correlations, depending on the spin of dark matter. We have shown the differential event rates for spin-independent DM-nucleon scattering at current direct detection experiments for fermion and scalar dark matter in detail. We have imposed the bounds from direct detection, relic density condition as well as LHC dijet searches to constrain the parameter space for the mass and couplings of the spin-2 mediator for both weak-scale and light dark matter cases.

\section*{Acknowledgments}

We would like to thank Deog-Ki Hong and Jure Zupan for helpful discussion on nucleon form factors. 
The work of YJK and HML is supported in part by Basic Science Research Program through the National Research Foundation of Korea (NRF) funded by the Ministry of Education, Science and Technology (NRF-2016R1A2B4008759).  The work of YJK was supported by IBS under the project code, IBS-R018-D1.
ACM is supported by the Mexican National Council for Science and Technology (CONACyT) scholarship scheme.

\def\theequation{A.\arabic{equation}}

\setcounter{equation}{0}

\vskip0.8cm
\noindent
{\Large \bf Appendix A: Differential scattering event rate} 
\vskip0.4cm
\noindent

The differential event rate per unit time per unit recoil energy for DM-nucleon elastic scattering is given \cite{fitz} by
\bea
\frac{dR}{dE_R}= \left\langle \frac{\rho_\chi m_T}{\mu^2_T m_\chi v}  \frac{d\sigma}{d\cos\theta}\right\rangle
\eea
where $\rho_\chi$ is the DM number density in the solar system, $m_T, \mu_T$ are the nucleus mass and the reduced mass of the DM-nucleus system, respectively, $v$ is the relative velocity between dark matter and nucleus, and $\frac{d\sigma}{d\cos\theta}$ is the differential scattering cross section with respect to the cosine of the scattering angle $\theta$ in the center of mass frame, given by
\bea
\frac{d\sigma}{d\cos\theta}= \frac{1}{2j_\chi+1} \frac{1}{2j+1} \sum_s \frac{1}{32\pi} \frac{|{\cal M}|^2}{(m_\chi+m_T)^2}
\eea 
where $j_\chi, j$ are the spins of dark matter and nucleus, respectively, and ${\cal M}$ is the scattering amplitude. 
Here, we note that $\langle\,\,\rangle$ is the average over the halo velocity distribution, namely, $\int_{\rm v_{\rm min}=q/(2\mu_T)} d^3 v f(v)$ where $f(v)$ is the velocity distribution function and  $v_{\rm min}$ is the minimum relative velocity to make the nuclear recoil happen for a given momentum transfer $q$.  

In reality, the scattering event rate at experiments depends on the detector material and mass.
The event rate per unit time per unit recoil energy per detector mass is given \cite{fitz} by
\bea
\frac{dR_D}{dE_R}= N_T\cdot \frac{\rho_\chi m_T}{32\pi m^3_\chi m^2_N} \cdot \left\langle  \frac{1}{v}\sum_{i,j}\sum_{N,N'=n,p} c^{(N)}_i c^{(N')}_j F_{ij}^{(N,N')}(v^2,q^2) \right\rangle
\eea
where $N_T$ is the number of target nuclei in a given detector,  $c^{(N)}_i$ are the coefficients of non-relativistic nucleon operators, ${\cal O}^{\rm NR}_i$, in the effective Lagrangian, and $F_{ij}^{(N,N')}(v^2,q^2) $ are the nucleon form factors, with symmetric property under $(i,N)\leftrightarrow (j,N')$.

\def\theequation{B.\arabic{equation}}

\setcounter{equation}{0}

\vskip0.8cm
\noindent
{\Large \bf Appendix B: DM-nucleon scattering amplitudes for dark matter} 
\vskip0.4cm
\noindent

In this appendix, we present the details on the calculation of nucleon scattering amplitudes with spin-2 mediator, in the order of fermion, scalar and vector dark matter.

\section*{Fermion dark matter}

Fom eqs.~(\ref{fdm-tensor}), the traceless part of the effective interactions for fermion dark matter becomes
\bea
16F_T(q^2){\tilde T}^\chi_{\mu\nu} {\tilde T}^{N,\mu\nu}
&=&F_T\Big[ (2(p_1+p_2)\cdot (k_1+k_2)) ({\bar u}_\chi(k_2)\gamma_\mu u_\chi(k_1)) ({\bar u}_N(p_2)\gamma^\mu u_N(p_1)) \nonumber \\
&&- ({\bar u}_\chi(k_2) (\slashed{k}_1+\slashed{k}_2) u_\chi(k_1))({\bar u}_N(p_2) (\slashed{p}_1+\slashed{p}_2) u_N(p_1))  \nonumber \\
&&+ 2 ({\bar u}_\chi(k_2) (\slashed{p}_1+\slashed{p}_2) u_\chi(k_1))({\bar u}_N(p_2) (\slashed{k}_1+\slashed{k}_2) u_N(p_1)) \Big] \nonumber \\
&=&F_T\Big[ (2(p_1+p_2)\cdot (k_1+k_2)) ({\bar u}_\chi(k_2)\gamma_\mu u_\chi(k_1)) ({\bar u}_N(p_2)\gamma^\mu u_N(p_1)) \nonumber \\
&&-4m_\chi m_N ({\bar u}_\chi(k_2) u_\chi(k_1))({\bar u}_N(p_2) u_N(p_1))  \nonumber \\
&&+ 2 ({\bar u}_\chi(k_2) (\slashed{p}_1+\slashed{p}_2) u_\chi(k_1))({\bar u}_N(p_2) (\slashed{k}_1+\slashed{k}_2) u_N(p_1)) \Big]  \label{tracefree}
\eea
where use is made of Dirac equations, $\slashed{p}u_N(p)=m_N u_N(p)  $ and ${\bar u}_N(p)\slashed{p}= {\bar u}_N(p) m_N$, etc. 
Using Gordon identities,
\bea
{\bar u}_\chi (k_2) \gamma^\mu u_\chi(k_1)&=&\frac{1}{2m_\chi} {\bar u}_\chi(k_2) \Big((k_1+k_2)^\mu-i\sigma^{\mu\rho}q_\rho \Big)u_\chi(k_1),   \label{GI1} \\
{\bar u}_N (p_2) \gamma^\nu u_N(p_1)&=&\frac{1}{2m_N} {\bar u}_N(p_2) \Big((p_1+p_2)^\nu+i\sigma^{\nu\lambda}q_\lambda \Big) u_N(p_1),  \label{GI2}
\eea
we can rewrite the vector operators in terms of scalar and tensor operators and obtain
\bea
16F_T(q^2){\tilde T}^\chi_{\mu\nu} {\tilde T}^{N,\mu\nu}&=&F_T\bigg[ \frac{(P\cdot K)}{2m_\chi m_N}\, \bigg((K\cdot P)({\bar u}_\chi(k_2) u_\chi(k_1))({\bar u}_N(p_2)  u_N(p_1)) \nonumber \\
&&+ ({\bar u}_\chi(k_2) u_\chi(k_1))(K_\nu {\bar u}_N(p_2)i\sigma^{\nu\lambda} q_\lambda u_N(p_1))  \nonumber \\
&&-  (P_\mu{\bar u}_\chi(k_2)i\sigma^{\mu\rho}q_\rho u_\chi(k_1))({\bar u}_N(p_2)  u_N(p_1)) \nonumber \\
&&-( {\bar u}_\chi(k_2)i\sigma^{\mu\rho}q_\rho u_\chi(k_1))({\bar u}_N(p_2)i\sigma_{\mu\lambda} q^\lambda u_N(p_1))  \bigg)  \nonumber \\
&&- 4 m_\chi m_N ({\bar u}_\chi(k_2) u_\chi(k_1))({\bar u}_N(p_2)  u_N(p_1))  \nonumber \\
&&+\frac{1}{2m_\chi m_N} \bigg((K\cdot P)^2({\bar u}_\chi(k_2) u_\chi(k_1))({\bar u}_N(p_2)  u_N(p_1)) \nonumber \\
&&-(P_\mu {\bar u}_\chi(k_2)i\sigma^{\mu\rho}q_\rho u_\chi(k_1))(K_\nu {\bar u}_N(p_2)i\sigma^{\nu\lambda} q_\lambda u_N(p_1)) \nonumber \\
&&+(K\cdot P)({\bar u}_\chi(k_2) u_\chi(k_1))(K_\nu {\bar u}_N(p_2)i\sigma^{\nu\lambda} q_\lambda u_N(p_1)) \nonumber \\
&&-(K\cdot P)  (P_\mu {\bar u}_\chi(k_2)i\sigma^{\mu\rho}q_\rho u_\chi(k_1))({\bar u}_N(p_2)  u_N(p_1))\bigg)  \bigg]
\label{tracefree}
\eea
where $P^\mu\equiv (p_1+p_2)^\mu$, $K^\mu\equiv (k_1+k_2)^\mu$ and $q^\mu\equiv (k_1-k_2)^\mu=(p_2-p_1)^\mu$. 
Using $2p_1\cdot k_1=s-m^2_N-m^2_\chi=2p_2\cdot k_2$ and $2p_1\cdot k_2=-u+m^2_N+m^2_\chi=2 p_2\cdot k_1$ for nucleon momenta, we note the approximate formula,
\bea
P\cdot K=(p_1+p_2)\cdot (k_1+k_2)= s-u \simeq 4 m_\chi m_N.
\label{approx}
\eea
where use is made of  $s\simeq (m_\chi+m_N)^2$ and $u\simeq (m_\chi-m_N)^2$ in the non-relativistic limit. 
The above nucleon operators can be matched to non-relativistic nucleon operators as in Ref.~\cite{fitz}, with the exception, the operator in the 7th line in eq.~(\ref{tracefree}), which is suppressed for a small momentum transfer as will be shown later.

Consequently, from eq.~(\ref{scatt-fermion0}) with eqs.~(\ref{tracefree}) and (\ref{trace}), 
 we get the scattering amplitude between fermion dark matter and nucleon as follows,
 \bea
 {\cal M}_\chi
 &=& \frac{ic_\chi c_\psi}{2m^2_G \Lambda^2} \, \bigg\{ \frac{1}{8}F_T\bigg[\frac{(P\cdot K)}{2m_\chi m_N}\, \bigg((K\cdot P)({\bar u}_\chi(k_2) u_\chi(k_1))({\bar u}_N(p_2)  u_N(p_1)) \nonumber \\
&&+ ({\bar u}_\chi(k_2) u_\chi(k_1))(K_\nu {\bar u}_N(p_2)i\sigma^{\nu\lambda} q_\lambda u_N(p_1))  \nonumber \\
&&-  (P_\mu{\bar u}_\chi(k_2)i\sigma^{\mu\rho}q_\rho u_\chi(k_1))({\bar u}_N(p_2)  u_N(p_1)) \nonumber \\
&&-( {\bar u}_\chi(k_2)i\sigma^{\mu\rho}q_\rho u_\chi(k_1))({\bar u}_N(p_2)i\sigma_{\mu\lambda} q^\lambda u_N(p_1))  \bigg)  \nonumber \\
&&- 4 m_\chi m_N ({\bar u}_\chi(k_2) u_\chi(k_1))({\bar u}_N(p_2)  u_N(p_1))  \nonumber \\
&&+\frac{1}{2m_\chi m_N} \bigg((K\cdot P)^2({\bar u}_\chi(k_2) u_\chi(k_1))({\bar u}_N(p_2)  u_N(p_1)) \nonumber \\
&&-(P_\mu {\bar u}_\chi(k_2)i\sigma^{\mu\rho}q_\rho u_\chi(k_1))(K_\nu {\bar u}_N(p_2)i\sigma^{\nu\lambda} q_\lambda u_N(p_1)) \nonumber \\
&&+(K\cdot P)({\bar u}_\chi(k_2) u_\chi(k_1))(K_\nu {\bar u}_N(p_2)i\sigma^{\nu\lambda} q_\lambda u_N(p_1)) \nonumber \\
&&-(K\cdot P)  (P_\mu {\bar u}_\chi(k_2)i\sigma^{\mu\rho}q_\rho u_\chi(k_1))({\bar u}_N(p_2)  u_N(p_1)) \bigg) \bigg] \nonumber \\
&&-\frac{1}{6} m_\chi m_N F_S ({\bar u}_\chi(k_2)u_\chi(k_1))({\bar u}_N(p_2)u_N(p_1))
  \bigg\}. \label{scatt-fermion}
 \eea 
 
As a result, from the scattering amplitude at the nucleon level given in eq.~(\ref{scatt-fermion}), the effective interactions between fermion dark matter and nucleons are given by
\bea
\mathcal{L}_{\chi,\rm eff}&=&\frac{c_\chi c_\psi}{2m_G^2 \Lambda^2}\bigg[\bigg\{F_T \bigg( \frac{1}{2}(P\cdot K)^2 +\frac{m_\chi}{2m_N}(P\cdot K){\vec q}^2 +\frac{m_N}{2m_\chi}(P\cdot K){\vec q}^2-2m_{\chi}^2 m_N^2+\frac{1}{4}{\vec q}^4 \bigg) \nonumber \\
&&-\frac{2}{3}F_S m_\chi^2 m_N^2 \bigg\}  {\mathcal{O}}_1^{\rm NR} -F_T m_N\Big(2m_\chi  (P\cdot K)+m_N {\vec q}^2\Big) {\mathcal{O}}_3^{\rm NR} -F_T(P\cdot K){\vec q}^2  {\mathcal{O}}_4^{\rm NR}  \nonumber \\
&&-F_T m_N \Big( 2m_N(P\cdot K)+m_\chi {\vec q}^2 \Big) {\mathcal{O}}_5^{\rm NR}+F_T m_N^2 (P\cdot K) {\mathcal{O}}_6^{\rm NR}+4F_Tm_N ^3 m_\chi  {\mathcal{O}}_3^{\rm NR} {\mathcal{O}}_5^{\rm NR} \bigg].\nonumber \\   \label{feff}
\eea
Then, using eq.~(\ref{feff}) with eq.~(\ref{approx}), we obtain the effective Lagrangian between fermion dark matter and nucleons as follows,
\bea
\mathcal{L}_{\chi,\rm eff}&=&\frac{c_\chi c_\psi}{2m_G^2 \Lambda^2}\bigg[\bigg\{F_T \bigg(6 m_{\chi}^2 m_N^2 + 2 (m_{\chi}^2+m_N^2 ){\vec q}^2 +\frac{{\vec q}^4}{4} \bigg)-\frac{2}{3}F_S m_\chi^2 m_N^2 \bigg\}  {\mathcal{O}}_1^{\rm NR}  \nonumber \\
&&-F_T m_N^2\Big(8 m_{\chi}^2 +{\vec q}^2\Big) {\mathcal{O}}_3^{\rm NR} -4 m_\chi m_N F_T\, {\vec q}^2  {\mathcal{O}}_4^{\rm NR} -F_T m_N m_\chi \Big( 8 m_N^2+{\vec q}^2 \Big) {\mathcal{O}}_5^{\rm NR} \nonumber \\
&&+4 m_\chi m_N^3 F_T  {\mathcal{O}}_6^{\rm NR}+4 F_T m_N^3 m_\chi  {\mathcal{O}}_3^{\rm NR} {\mathcal{O}}_5^{\rm NR} \bigg].  \label{feff2}
\eea
Here, we note that a factor $\int \frac{d^3p}{(2\pi)^3\sqrt{2E}} \, a^{(\dagger)}_N$ per each nucleon state or $\int \frac{d^3p}{(2\pi)^3\sqrt{2E}} \, a^{(\dagger)}_\chi$ per each dark matter state, with dimension $E$, are to be multiplied as overall factors such that the above effective Lagrangian for nucleons has a dimension 4.

 \section*{Scalar dark matter}

From eq.~(\ref{sdm-tensor}), the traceless part of the effective interactions for scalar dark matter is given by
\bea
4F_T(q^2) {\tilde T}^S_{\mu\nu} {\tilde T}^{N,\mu\nu}&=&F_T\Big(2{\bar u}_N(p_2)(\slashed{k}_1 k_2\cdot(p_1+p_2)+\slashed{k}_2 k_1\cdot (p_1+p_2)) u_N(p_1) \nonumber \\
&&-2 m_N (k_1\cdot k_2)({\bar u}_N(p_2) u_N(p_1)) \Big)
\eea
Then, using $k_1\cdot (p_1+p_2)=k_2\cdot(p_1+p_2)=(s-u)/2$ and Gordon identity, we can rewrite the above result as 
\bea
4F_T(q^2) {\tilde T}^S_{\mu\nu} {\tilde T}^{N,\mu\nu}&=&F_T\Big[ \frac{(P\cdot K)}{2m_N}\,\bigg(  (P\cdot K) ({\bar u}_N(p_2) u_N(p_1))+K_\nu {\bar u}_N(p_2)i\sigma^{\nu\lambda}q_\lambda u_N(p_1) \bigg) \nonumber \\
&& -2 m_N (k_1\cdot k_2)( {\bar u}_N(p_2) u_N(p_1)) \Big].   \label{tracefreeS}
\eea

As a result, from eq.~(\ref{scatt-scalar0}) with eqs.~(\ref{tracefreeS}) and (\ref{traceS}), the scattering amplitude between scalar dark matter and nucleon as follows,
 \bea
{\cal M}_S
&=& \frac{ic_S c_\psi}{2m^2_G \Lambda^2} \, \bigg[F_T\bigg( \frac{(P\cdot K)}{4m_N}\,\bigg(  (P\cdot K) ({\bar u}_N(p_2) u_N(p_1))+(K_\nu {\bar u}_N(p_2)i\sigma^{\nu\lambda}q_\lambda u_N(p_1)) \bigg) \nonumber \\
&&- m_N (k_1\cdot k_2)( {\bar u}_N(p_2) u_N(p_1)) \bigg) -\frac{1}{3} m_N F_S (2m^2_S-k_1\cdot k_2)\, ({\bar u}_N(p_2)u_N(p_1))
\bigg].  \label{scatt-scalar}
 \eea
Here, we note that the tensor operator ${\bar N}i\sigma^{\nu\lambda}q_\lambda N$ can be written as the sum of vector and scalar operators by Gordon identity. 

Consequently, from eq.~(\ref{scatt-scalar}), we obtain the effective Lagrangian for scalar dark matter as
\bea
\mathcal{L}_{S, \rm eff}&=&\frac{c_S c_\psi}{2m_G^2\Lambda^2}\bigg[ 2F_T\bigg( m_S m_N(P\cdot K)-m_N^2(k_1\cdot k_2) \bigg) -\frac{2}{3}F_Sm_N^2(2m_S^2-k_1\cdot k_2)  \bigg] {\cal O}_1^{\rm NR} \nonumber \\  \label{seff}
\eea

 \section*{Vector dark matter}

The energy-momentum tensor for a vector DM $X$  is, in momentum space,
\bea
T^X_{\mu\nu}=-\Big(m^2_X C_{\mu\nu,\alpha\beta}+W_{\mu\nu,\alpha\beta} \Big) \epsilon^{\alpha}(k_1)\epsilon^{*\beta}(k_2)
\eea
where $\epsilon^\alpha(k)$ is the polarization vector for the vector DM and 
\bea
W_{\mu\nu,\alpha\beta}&\equiv & -\eta_{\alpha\beta} k_{1\mu} k_{2\nu} -\eta_{\mu\alpha} (k_1\cdot k_2\,\eta_{\nu\beta}-k_{1\beta} k_{2\nu})+\eta_{\mu\beta} k_{1\nu} k_{2\alpha} \nonumber \\
&& -\frac{1}{2}\eta_{\mu\nu}(k_{1\beta}k_{2\alpha}-k_1\cdot k_2\, \eta_{\alpha\beta}) +(\mu\leftrightarrow\nu ).
\eea
Likewise as before, the vector DM  is incoming into the vertex with momentum $k_1$ and is outgoing from the vertex with momentum $k_2$.
Then, the trace of the energy-momentum tensor is given by
\bea
T^X=2 m^2_X \eta_{\alpha\beta} \epsilon^{\alpha}(k_1)\epsilon^{*\beta}(k_2).  \label{vdm-scalar}
\eea
\bea
{\tilde T}^X_{\mu\nu}=-\Big(m^2_X C_{\mu\nu,\alpha\beta}+W_{\mu\nu,\alpha\beta}+\frac{1}{2}m^2_X \eta_{\mu\nu}\eta_{\alpha\beta} \Big) \epsilon^{\alpha}(k_1)\epsilon^{*\beta}(k_2).\label{vdm-tensor}
\eea
Here, we note that $W_{\mu\nu,\alpha\beta} \eta^{\mu\nu}=0$, due to the fact that the energy-momentum tensor for transverse polarizations of vector dark matter is trace-free.

We consider the elastic scattering between the vector DM and the nucleon, $X(k_1)+N(p_1)\rightarrow X(k_2)+N(p_2)$. 
First, with eq.~(\ref{nucl-tensor}), we note  the following Lorentz contractions,
$$
-4m^2_X C_{\mu\nu,\alpha\beta} \langle N(p_2)|{\tilde T}^{\psi,\mu\nu} |N(p_1)\rangle
$$
\bea
&=& m^2_X F_T\Big( 2P_\beta {\bar u}_N(p_2)\gamma_\alpha u_N(p_1) + 2P_\alpha {\bar u}_N(p_2)\gamma_\beta u_N(p_1)  \nonumber \\ 
&&-2m_N \eta_{\alpha\beta} {\bar u}_N(p_2) u_N(p_1) \Big),
\eea
$$
-4 W_{\mu\nu,\alpha\beta}\langle N(p_2)|{\tilde T}^{\psi,\mu\nu}|N(p_1)\rangle 
$$
\bea
&=&F_T\Big[ \eta_{\alpha\beta} \Big(-(K\cdot P){\bar u}_N(p_2) (\slashed{k}_1+\slashed{k}_2) u_N(p_1)+ 4m_N (k_1\cdot k_2){\bar u}_N(p_2) u_N(p_1)   \Big) \nonumber \\
&&-2(k_1\cdot k_2) \Big(P_\beta {\bar u}_N(p_2)\gamma_\alpha u_N(p_1) + P_\alpha {\bar u}_N(p_2)\gamma_\beta u_N(p_1)  \Big) \nonumber \\
&&+(K\cdot P) \Big(k_{1\beta} {\bar u}_N(p_2)\gamma_\alpha u_N(p_1)+ k_{2\alpha} {\bar u}_N(p_2)\gamma_\beta u_N(p_1)\Big) \nonumber \\
&& +2k_{1\beta}P_\alpha {\bar u}_N(p_2) \slashed{k}_2 u_N(p_1)+ 2k_{2\alpha}P_\beta {\bar u}_N(p_2) \slashed{k}_1 u_N(p_1) \nonumber \\
&&-4m_N k_{1\beta} k_{2\alpha} {\bar u}_N(p_2)  u_N(p_1) \Big], \\
-2 m^2_X \eta_{\mu\nu}\eta_{\alpha\beta}{\tilde T}^{N,\mu\nu}  &=& 0.
\eea
Then, from the above results with eq.~(\ref{vdm-tensor}), the effective interactions for traceless parts are 
$$
4{\tilde T}^X_{\mu\nu}\langle N(p_2)| {\tilde T}^{\psi,\mu\nu}|N(p_1)\rangle 
$$
\bea
&=&-4\epsilon^{\alpha}(k_1)\epsilon^{*\beta}(k_2)   \Big(m^2_X C_{\mu\nu,\alpha\beta} +W_{\mu\nu,\alpha\beta}+\frac{1}{2} m^2_X \eta_{\mu\nu}\eta_{\alpha\beta}\Big)\langle N(p_2)|{\tilde T}^{\psi,\mu\nu}|N(p_1)\rangle   \nonumber \\
&=& \epsilon^{\alpha}(k_1)\epsilon^{*\beta}(k_2) F_T \bigg[2m^2_X\Big(P_\beta {\bar u}_N(p_2)\gamma_\alpha u_N(p_1) + P_\alpha {\bar u}_N(p_2)\gamma_\beta u_N(p_1) \nonumber \\
&&-m_\psi \eta_{\alpha\beta} {\bar u}_N(p_2)  u_N(p_1)\Big)   \nonumber \\
&&+\eta_{\alpha\beta}\Big(-(K\cdot P){\bar u}_N(p_2) \slashed{K} u_N(p_1)+ 4m_N (k_1\cdot k_2){\bar u}_N(p_2)  u_N(p_1)\Big)   \nonumber \\
&&-2(k_1\cdot k_2) \Big(P_\beta {\bar u}_N(p_2)\gamma_\alpha u_N(p_1) + P_\alpha {\bar u}_N(p_2)\gamma_\beta u_N(p_1)  \Big) \nonumber \\
&& +(K\cdot P) \Big(k_{1\beta} {\bar u}_N(p_2)\gamma_\alpha u_N(p_1)+ k_{2\alpha} {\bar u}_N(p_2)\gamma_\beta u_N(p_1)\Big) \nonumber \\
&& +2k_{1\beta}P_\alpha {\bar u}_N(p_2) \slashed{k}_2 u_N(p_1)+ 2k_{2\alpha}P_\beta {\bar u}_N(p_2) \slashed{k}_1 u_N(p_1) \nonumber \\
&&-4m_N k_{1\beta} k_{2\alpha} {\bar u}_N(p_2)  u_N(p_1) \bigg].
\eea
After using the Gordon identities, (\ref{GI1}) and (\ref{GI2}) to rewrite the vector operators in terms of scalar and tensor operators,  we obtain
$$
4{\tilde T}^X_{\mu\nu}  \langle N(p_2)|{\tilde T}^{\psi,\mu\nu}|N(p_1)\rangle
$$
\bea
&=& \epsilon^{\alpha}(k_1)\epsilon^{*\beta}(k_2) F_T\bigg[\frac{m_X^2}{m_N}\Big(2P_\alpha  P_\beta {\bar u}_N(p_2) u_N(p_1) + P_\alpha {\bar u}_N(p_2)i\sigma_{\beta \lambda}q^\lambda u_N(p_1)  \nonumber \\
&&+ P_\beta {\bar u}_N(p_2)i\sigma_{\alpha \lambda}q^\lambda u_N(p_1)\Big) -2m_X^2 m_N \eta_{\alpha\beta} {\bar u}_N(p_2)  u_N(p_1)   \nonumber \\
&&-\frac{(K\cdot P)}{2m_N}\eta_{\alpha\beta}\Big((K\cdot P){\bar u}_N(p_2)  u_N(p_1)+K_\nu{\bar u}_N(p_2) i\sigma^{\nu \lambda}q_\lambda  u_N(p_1) \Big)  \nonumber \\
&&+ 4m_N \eta_{\alpha \beta}(k_1\cdot k_2){\bar u}_N(p_2)  u_N(p_1)   \nonumber \\
&&-\frac{(k_1\cdot k_2)}{m_N} \Big(2P_\alpha  P_\beta {\bar u}_N(p_2) u_N(p_1) + P_\alpha {\bar u}_N(p_2)i\sigma_{\beta \lambda}q^\lambda u_N(p_1)  \nonumber \\
&&+ P_\beta {\bar u}_N(p_2)i\sigma_{\alpha \lambda}q^\lambda u_N(p_1)  \Big) \nonumber \\
&& +\frac{(K\cdot P)}{2m_N} \Big(k_{1\beta} P_\alpha {\bar u}_N(p_2) u_N(p_1)+k_{2\alpha} P_\beta {\bar u}_N(p_2) u_N(p_1) \nonumber \\ 
&&+ k_{1\beta} {\bar u}_N(p_2)i\sigma_{\alpha \lambda}q^\lambda u_N(p_1) + k_{2\alpha} {\bar u}_N(p_2)i\sigma_{\beta \lambda}q^\lambda u_N(p_1)\Big) \nonumber \\
&&+\frac{(K\cdot P)}{2m_N}\Big( k_{1\beta}P_\alpha {\bar u}_N(p_2) u_N(p_1) +k_{2\alpha}P_\beta {\bar u}_N(p_2)u_N(p_1)  \Big) \nonumber \\
&&+\frac{1}{m_N}\Big( k_{1\beta}P_\alpha k_{2\nu}  {\bar u}_N(p_2) i\sigma^{\nu \lambda}q_\lambda u_N(p_1) + k_{2\alpha}P_\beta k_{1\nu}  {\bar u}_N(p_2) i\sigma^{\nu \lambda}q_\lambda u_N(p_1)  \Big) \nonumber \\
&&-4m_N k_{1\beta} k_{2\alpha} {\bar u}_N(p_2)  u_N(p_1) \bigg]. 
\eea

On the other hand, from eqs.~(\ref{vdm-scalar}) and (\ref{nucl-trace}), the effective interactions for trace parts are
\bea
4T^X \langle N(p_2)|T^\psi|N(p_1)\rangle=-8 m^2_X m_N F_S  (\eta_{\alpha\beta} \epsilon^{\alpha}(k_1)\epsilon^{*\beta}(k_2) )  ({\bar u}_N(p_2)u_N(p_1)). 
\eea

Consequently, 
 we get the scattering amplitude between vector dark matter and nucleon, as follows,
 \bea
{\cal M}_X&=& \frac{ic_X c_\psi}{2m^2_G \Lambda^2} \, \langle N(p_2)|\left(2{\tilde T}^X_{\mu\nu} {\tilde T}^{\psi,\mu\nu} -\frac{1}{6} {T}^X{ T}^\psi \right)|N(p_1)\rangle \nonumber \\
&=&\frac{ic_X c_\psi}{2m_G^2\Lambda^2}  \epsilon^{\alpha}(k_1)\epsilon^{*\beta}(k_2) \bigg\{F_T\bigg[\frac{m_X^2}{2m_N}\Big(2P_\alpha  P_\beta {\bar u}_N(p_2) u_N(p_1) + P_\alpha {\bar u}_N(p_2)i\sigma_{\beta \lambda}q^\lambda u_N(p_1)  \nonumber \\
&&+ P_\beta {\bar u}_N(p_2)i\sigma_{\alpha \lambda}q^\lambda u_N(p_1)\Big) -m_X^2 m_N \eta_{\alpha\beta} {\bar u}_N(p_2)  u_N(p_1)   \nonumber \\
&&-\frac{(K\cdot P)}{4m_N}\eta_{\alpha\beta}\Big((K\cdot P){\bar u}_N(p_2)  u_N(p_1)+K_\nu{\bar u}_N(p_2) i\sigma^{\nu \lambda}q_\lambda  u_N(p_1) \Big)  \nonumber \\
&&+ 2m_N \eta_{\alpha \beta}(k_1\cdot k_2){\bar u}_N(p_2)  u_N(p_1)   \nonumber \\
&&-\frac{(k_1\cdot k_2)}{2m_N} \Big(2P_\alpha  P_\beta {\bar u}_N(p_2) u_N(p_1) + P_\alpha {\bar u}_N(p_2)i\sigma_{\beta \lambda}q^\lambda u_N(p_1)  \nonumber \\
&&+ P_\beta {\bar u}_N(p_2)i\sigma_{\alpha \lambda}q^\lambda u_N(p_1)  \Big) \nonumber \\
&& +\frac{(K\cdot P)}{4m_N} \Big(2k_{1\beta} P_\alpha {\bar u}_N(p_2) u_N(p_1)+2k_{2\alpha} P_\beta {\bar u}_N(p_2) u_N(p_1) \nonumber \\ 
&&+ k_{1\beta} {\bar u}_N(p_2)i\sigma_{\alpha \lambda}q^\lambda u_N(p_1) + k_{2\alpha} {\bar u}_N(p_2)i\sigma_{\beta \lambda}q^\lambda u_N(p_1)\Big) \nonumber \\
&&+\frac{1}{2m_N}\Big( k_{1\beta}P_\alpha k_{2\nu}  {\bar u}_N(p_2) i\sigma^{\nu \lambda}q_\lambda u_N(p_1) + k_{2\alpha}P_\beta k_{1\nu}  {\bar u}_N(p_2) i\sigma^{\nu \lambda}q_\lambda u_N(p_1)  \Big) \nonumber \\
&&-2m_N k_{1\beta} k_{2\alpha} {\bar u}_N(p_2)  u_N(p_1) \bigg)\bigg] \nonumber \\
&&+\frac{1}{3} m^2_X m_N F_S \eta_{\alpha\beta} ({\bar u}_N(p_2)u_N(p_1))
\bigg\}. \label{vector-scatt}
 \eea

The effective operators between vector dark matter and nucleon match with relativistic operators, as follows,
\bea
{\bar N}N  &\rightarrow & 2m_N f(\epsilon_1,\epsilon^*_2){\cal O}^{\rm NR}_1,  \label{nucl1} \\
\epsilon^{\alpha}_{1,2} {\bar N}i\sigma_{\alpha\lambda} q^\lambda N &\rightarrow&4i m_N^2\Big( {\vec s}_N\cdot({\vec \epsilon}_{1,2}\times \frac{\vec q}{m_N})\Big),   \label{nucl2}\\ 
k_{1,2\nu} {\bar N}i\sigma^{\nu\lambda}q_\lambda N  &\rightarrow& m_\chi\Big({\vec q}^2{\cal O}^{\rm NR}_1-4 m_N^2 {\cal O}^{\rm NR}_3\Big). 
 \label{nucl3}
\eea
The above results for vector dark matter are also included in Table~\ref{tabla:sencilla}.
Here, we note that the longitudinal polarization vector is given by $\epsilon^\mu_L(k)=(|{\vec k}|,k^0\,{\vec k}/|{\vec k}|)/m$ with $\epsilon_L\cdot \epsilon_L=-1$, and two transverse polarization vectors satisfy $\epsilon_T\cdot \epsilon_T=-1$ and $\epsilon_T\cdot \epsilon_L=0$.  
Then, in the rest frame of vector dark matter, we can take $\epsilon^\mu_T=(0,1,0,0), (0,0,0,1)$ and $\epsilon^\mu_L=(0,0,0,1)$. 
We find that there are ${\cal O}^{\rm NR}_{1,3}$ operators as in the case with the fermion and scalar dark matter, but there is a new spin-dependent interaction, ${\vec s}_N\cdot({\vec \epsilon}_{1,2}\times{\vec q})$. 

As a result, from eq.~(\ref{vector-scatt}) with eqs. (\ref{nucl1})-(\ref{nucl3}), we obtain the effective Lagrangian for vector dark matter as
\bea
\mathcal{L}_{X,{\rm eff}}&=&\frac{c_X c_\psi}{2m_G^2 \Lambda^2}\bigg[ \bigg\{F_T\bigg( 2m_X^2 P_\alpha P_\beta -2m_X^2 m_N^2 \eta_{\alpha\beta}-\frac{1}{2}(K\cdot P)^2\eta_{\alpha \beta} \nonumber\\
&& -\frac{m_X}{2m_N}\vec{q}^2((K\cdot P)\eta_{\alpha\beta}-P_\alpha k_{1\beta}-k_{2\alpha}P_{\beta})+4m_N^2((k_1\cdot k_2)\eta_{\alpha \beta}-k_{2\alpha}k_{1\beta}) \nonumber \\
&&-2(k_1\cdot k_2)P_\alpha P_\beta  \bigg) +\frac{2}{3}F_S m_X^2 m_N^2 \eta_{\alpha\beta}  \bigg\}\epsilon^{\alpha} \epsilon^{* \beta}{\cal O}_{1}^{\rm NR}  
 \nonumber \\
&&+ 2F_T m_Xm_N \bigg( (K\cdot P)\eta_{\alpha\beta}-P_\alpha k_{1\beta}-k_{2\alpha}P_{\beta} \bigg)\epsilon^\alpha \epsilon^{*\beta}{\cal O}_3^{
\rm NR} \nonumber \\
&&+ F_T m_N \bigg(2m_X^2 P_\alpha -2(k_1\cdot k_2)P_\alpha+(K\cdot P)k_{2\alpha} \bigg)\epsilon^\alpha i\Big(\vec{s_N}\cdot(\vec{\epsilon}_2\times\frac{\vec{q}}{m_N})\Big) \nonumber \\
&&+ F_T m_N \bigg(2m_X^2 P_\beta -2(k_1\cdot k_2)P_\beta+(K\cdot P)k_{1\beta} \bigg)\epsilon^{*\beta} i\Big(\vec{s_N}\cdot(\vec{\epsilon}_1\times\frac{\vec{q}}{m_N})\Big) \bigg] \label{veff}
\eea
We note that for a zero momentum transfer, the above effective Lagrangian (\ref{veff}) is reduced to scalar operators $ {\mathcal{O}}_1^{\rm NR}$ only.

\def\theequation{C.\arabic{equation}}

\setcounter{equation}{0}

\vskip0.8cm
\noindent
{\Large \bf Appendix C: Twist-2 operators with zero momentum transfer} 
\vskip0.4cm
\noindent

For a small momentum transfer in the DM-nucleon scattering, we can alternatively use the nuclear matrix elements for twist-2 operators or trace parts of energy-momentum tensor for quarks and then obtain the scattering amplitudes between dark matter and nucleons. 

The traceless parts of energy-momentum tensor or twist-2 operators for quarks and gluons, ${\tilde T}^q_{\mu\nu}$ and ${\tilde T}^g_{\mu\nu}$, can be treated with speciality in the calculation of matrix elements for dark matter-nucleon scattering.
For a small momentum transfer, the spin-independent effective coupling can be derived from the following matrix elements between initial and final nucleon states with mass $m_N$~\cite{drees,hisano,solon}, 
\bea
\langle N| m_q{\bar q}q|N\rangle/m_N &\equiv & f^N_{Tq}\, {\bar u}_N(p) u_N(p),  \label{scalar-0}  \\
1-\sum_{u,d,s} f_{Tq} &\equiv & f_{TG}, \\
\langle N(p)| {\tilde T}^q_{\mu\nu}|N(p)\rangle&= & \frac{1}{m_N} \, \Big(p_\mu p_\nu -\frac{1}{4} m^2_N g_{\mu\nu}\Big)(q(2)+{\bar q}(2))\, {\bar u}_N(p) u_N(p),  \label{twistquark}  \\
\langle N(p)| {\tilde T}^g_{\mu\nu}|N(p)\rangle&=& \frac{1}{m_N} \, \Big(p_\mu p_\nu -\frac{1}{4} m^2_N g_{\mu\nu}\Big)G(2)\, {\bar u}_N(p) u_N(p)  \label{twistgluon}
\eea
where $q(2), {\bar q}(2)$ and $G(2)$ are the second moments of the parton distribution functions(PDFs) of quark, antiquark and gluon, respectively, 
\bea
q(2)+{\bar q}(2) &=& \int^1_0 dx\, x \,[q(x)+{\bar q}(x)],  \label{pdf2nd} \\
G(2)&=& \int^1_0 dx\, x\, g(x). 
\eea
Here, we note that $f^N_{Tq}$ and $f_{TG}$  denote the mass fractions of light quarks and gluons in a nucleon, respectively. 
The second moments of PDFs in a proton have scale dependence, so we evaluate them at  the scale $\mu=m_Z$ because the effective couplings are matched at the scale of the mediator particle \cite{hisano}.  

The mass fractions are  $f^p_{T_u}=0.023$, $f^p_{T_d}=0.032$ and $f^p_{T_s}=0.020$ for a proton and $f^n_{T_u}=0.017$, $f^n_{T_d}=0.041$ and $f^n_{T_s}=0.020$ for a neutron \cite{hisano}. On the other hand, the second moments of PDFs are calulated at the scale $\mu=m_Z$ using the CTEQ parton distribution as $G(2)=0.48$, $u(2)=0.22$, ${\bar u}(2)=0.034$, $d(2)=0.11$, ${\bar d}(2)=0.036$, $s(2)={\bar s}(2)=0.026$, $c(2)={\bar c}(2)=0.019$ and $b(2)={\bar b}(2)=0.012$ \cite{hisano}. 

On the other hand, the spin-dependent effective coupling is given by
\bea
a_N=\sum_{q=u,d,s} d_q \Delta q_N
\eea
where  $d_q$ is the effective coupling at the quark level and 
\bea
2s_\mu \Delta q_N\equiv \langle N| {\bar q} \gamma_\mu \gamma_5 q|N\rangle
\eea
with $s_\mu$ is the spin of a nucleon. 
Here, $\Delta q_N$ denotes spin fractions of light quarks in a nucleon and they are given by $\Delta u_p=0.77$, $\Delta d_p=-0.49$ and $\Delta s_p=-0.15$, for a proton \cite{hisano}.

As a result, we get the scattering amplitude between fermion dark matter and nucleon as follows,
\bea
{\cal M}_\chi &=& \frac{ic_\chi c_\psi}{2m^2_G \Lambda^2} \, \bigg[ 2{\tilde T}^\chi_{\mu\nu}\,\cdot \frac{1}{m_N} \, \Big(p_\mu p_\nu -\frac{1}{4} m^2_N g_{\mu\nu}\Big)(\psi(2)+{\bar \psi}(2))+\frac{1}{6} m_N f^N_{T\psi} {T}^\chi \bigg]  \nonumber \\
&=&  \frac{ic_\chi c_\psi}{2m^2_G \Lambda^2} \,  \bigg[ -\frac{1}{m_N}\,(\psi(2)+{\bar \psi}(2)) [p\cdot (k_1+k_2)] ({\bar u}_\chi(k_2) \slashed{p} u_\chi(k_1)) \nonumber \\
&& +\frac{1}{2} m_N m_\chi \Big( \psi(2)+{\bar \psi}(2)-\frac{1}{3} f^N_{T\psi} \Big) ({\bar u}_\chi(k_2) u_\chi(k_1))\bigg]{\bar u}_N(p) u_N(p). 
 \eea
Moreover, the scattering amplitudes between dark matter with other spins and nucleon can be obtained similarly, as follows,
\bea
{\cal M}_S &=& \frac{ic_S c_\psi}{2m^2_G \Lambda^2} \, \bigg[ 2{\tilde T}^S_{\mu\nu}\,\cdot \frac{1}{m_N} \, \Big(p_\mu p_\nu -\frac{1}{4} m^2_N g_{\mu\nu}\Big)(\psi(2)+{\bar \psi}(2))+\frac{1}{6} m_N f^N_{T\psi} {T}^S \bigg]  \nonumber \\
&=&\frac{ic_S c_\psi}{2m^2_G \Lambda^2} \, \bigg[ \frac{2}{m_N}\,(\psi(2)+{\bar \psi}(2)) \Big(\frac{1}{2}m^2_N (k_1\cdot k_2)-2(p\cdot k_1)(p\cdot k_2) \Big) \nonumber \\
&&-\frac{1}{3} m_N f^N_{T\psi} (2m^2_S - k_1\cdot k_2 ) \bigg]{\bar u}_N(p) u_N(p), 
\eea
and
\bea
{\cal M}_X &=& \frac{ic_X c_\psi}{2m^2_G \Lambda^2} \, \bigg[ 2{\tilde T}^X_{\mu\nu}\,\cdot \frac{1}{m_N} \, \Big(p_\mu p_\nu -\frac{1}{4} m^2_N g_{\mu\nu}\Big)(\psi(2)+{\bar \psi}(2))+\frac{1}{6} m_N f^N_{T\psi} {T}^X \bigg]  \nonumber \\
&=& \frac{ic_X c_\psi}{2m^2_G \Lambda^2} \,\epsilon^\alpha(k_1) \epsilon^{*\beta}(k_2) \nonumber \\
&&\quad\times \bigg[ \frac{2}{m_N}\bigg\{ 2p_\alpha p_\beta (k_1\cdot k_2-m^2_X) -\frac{1}{2} m^2_N\eta_{\alpha\beta} (2k_1\cdot k_2-m^2_X) + 2\eta_{\alpha\beta}(p\cdot k_1)(p\cdot k_2) \nonumber \\
&&\quad\quad\quad\quad\quad\quad +m^2_N k_{1\beta} k_{2\alpha}-2p_\alpha k_{1\beta} (p\cdot k_2) - 2p_\beta k_{2\alpha} (p\cdot k_1)  \bigg\} (\psi(2)+{\bar\psi}(2)) \nonumber \\
&&\quad\quad \quad + \frac{1}{3} m_N m^2_X f^N_{T\psi} \eta_{\alpha\beta} \bigg]{\bar u}_N(p) u_N(p).
\eea

\end{document}